\documentclass[acmtog]{acmart}

\usepackage{color}
\usepackage{ifthen}
\usepackage{float}
\usepackage{alltt}
\usepackage{amsmath}
\usepackage{amsthm}
\usepackage{newlfont}
\usepackage{floatflt}
\usepackage{wrapfig}
\usepackage{fixltx2e}
\usepackage{subfig}
\usepackage{multirow}
\usepackage{booktabs}
\usepackage{cleveref}
\usepackage{algorithmic}
\usepackage[export]{adjustbox}
\usepackage{mathtools, cuted}
\usepackage{CJKutf8}
\usepackage[ruled]{algorithm2e}

\setcitestyle{square}

\SetAlFnt{\small}
\SetAlCapFnt{\small}
\SetAlCapNameFnt{\small}
\SetAlCapHSkip{0pt}
\IncMargin{-\parindent}
\newcommand{\Caption}[2]{\caption[#1]{{\em #1} #2}}
\let\oldcaption\caption
\renewcommand{\caption}[2][]{\oldcaption[#1]{{\em #1} #2}}

\definecolor{figred}{rgb}{1,0,0}
\definecolor{figgreen}{rgb}{0,0.6,0}
\definecolor{figblue}{rgb}{0,0,1}
\definecolor{figpink}{rgb}{1,0.63,0.63}

\newcommand{\pseudocode}{Algorithm}
\floatstyle{plain}
\newfloat{algorithm}{tbhp}{lop}
\floatname{algorithm}{\pseudocode}

\newcommand{\filename}[1]{\url{#1}}
\newcommand{\foldername}[1]{\url{#1}}

\let\oldparagraph\paragraph
\iffalse

\renewcommand{\paragraph}[1]{\oldparagraph{\textbf{#1}.}}
\else

\renewcommand{\paragraph}[1]{\oldparagraph{{#1}.}}
\fi

\ifdefined\email
\else
\newcommand{\email}[1]{\url{#1}}
\fi

\newcommand{\forceCondition}{\textbf{FORCE}\xspace}
\newcommand{\positionCondition}{\textbf{POSITION}\xspace}
\acmJournal{TOG}
\acmYear{2022}
\acmVolume{41}
\acmNumber{6}
\acmArticle{268}
\acmMonth{12}
\acmDOI{10.1145/3550454.3555461}
\title[Force-Aware Interface via Electromyography for Natural VR/AR Interaction]{Force-Aware Interface via Electromyography for Natural VR/AR Interaction}

\author{Yunxiang Zhang}
\orcid{0000-0003-0189-1776}
\email{yunxiang.zhang@nyu.edu}
\affiliation{
 \institution{New York University}
 \country{USA}}
\affiliation{
 \institution{The Chinese University of Hong Kong}
 \country{Hong Kong SAR}}
 
\author{Benjamin Liang}
\email{ben.liang@nyu.edu}
\author{Boyuan Chen}
\email{boyuan.chen@nyu.edu}
\affiliation{
 \institution{New York University}
 \country{USA}}
 
\author{Paul Torrens}
\email{pt50@nyu.edu}
\author{S. Farokh Atashzar}
\email{sfa7@nyu.edu}
\affiliation{
 \institution{New York University}
 \country{USA}}
 
\author{Dahua Lin}
\email{dhlin@ie.cuhk.edu.hk}
\affiliation{
 \institution{The Chinese University of Hong Kong}
 \country{Hong Kong SAR}}
\affiliation{
 \institution{Shanghai Artificial Intelligence Laboratory}
 \country{China}}
 
\author{Qi Sun}
\orcid{0000-0002-3094-5844}
\email{qisun@nyu.edu}
\affiliation{
 \institution{New York University}
 \country{USA}}
\begin{teaserfigure}
\centering
\subfloat[The reaction of virtual balls with different physical properties to varying level of pressing force decoded from forearm electromyography signals.]{
    \label{fig:teaser-top}
    \includegraphics[width=0.99\linewidth]{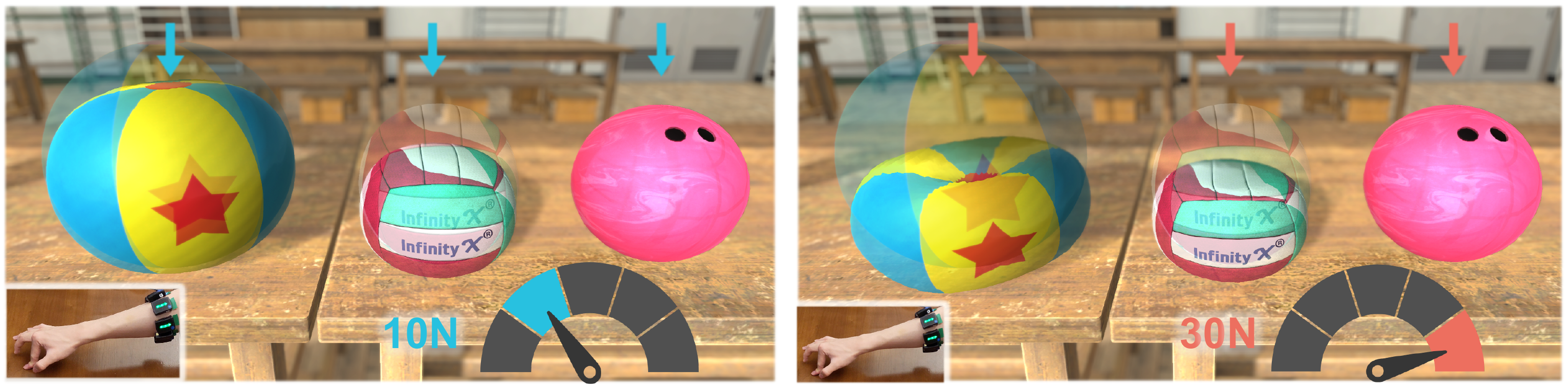}
  } \\
\subfloat[Our system provides a natural and intuitive interface for capturing user-generated forces and letting them take effects in the virtual environment.]{
    \label{fig:teaser-bottom}
    \includegraphics[width=0.99\linewidth]{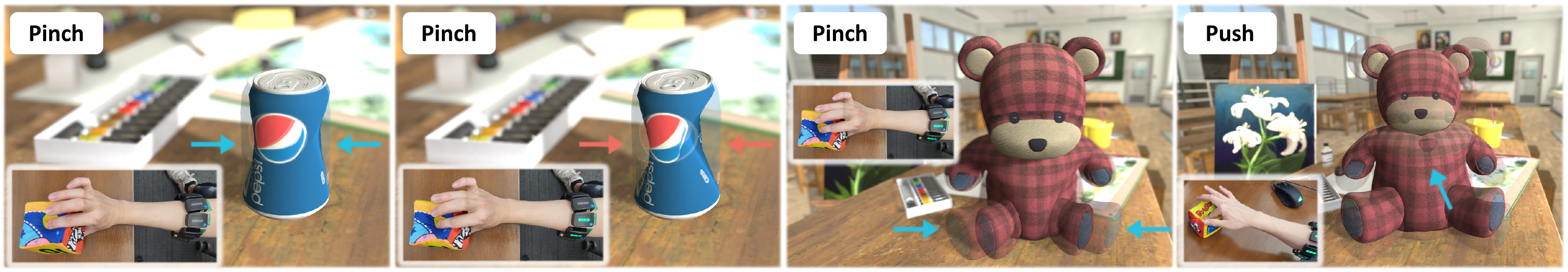}
  }
\Caption{Electromyography (EMG)-based neural interface with learned muscular force decoder.}
{With user-generated physical forces being decoded by our interface and applied to virtual objects in real-time,~\subref{fig:teaser-top} illustrates the deformation of a beach ball, a volleyball, and a bowling ball in VR, subject to pressing force of varying intensities. The beach ball is softer than the volleyball and thus exhibits larger deformation under the same force level, while the bowling ball is rigid and barely deforms within the force range of finger pressing. Our scheme helps users better perceive/distinguish the physical properties of virtual objects, in a way similar to how they approach it in the real world. \subref{fig:teaser-bottom} shows force-enabled virtual interactions via our system, with enhanced physical realism. 3D asset credits to SbbUtutuya, Virtual Method, TGameAssets at Unity, and TankStorm at Sketchfab.}
\label{fig:teaser}
\end{teaserfigure}

\setcopyright{acmcopyright}
\begin{abstract}
While tremendous advances in visual and auditory realism have been made for virtual and augmented reality (VR/AR), introducing a plausible sense of physicality into the virtual world remains challenging. Closing the gap between real-world physicality and immersive virtual experience requires a closed interaction loop: applying user-exerted physical forces to the virtual environment and generating haptic sensations back to the users. However, existing VR/AR solutions either completely ignore the force inputs from the users or rely on obtrusive sensing devices that compromise user experience.

By identifying users' muscle activation patterns while engaging in VR/AR, we design a learning-based neural interface for natural and intuitive force inputs. Specifically, we show that lightweight electromyography sensors, resting non-invasively on users' forearm skin, inform and establish a robust understanding of their complex hand activities. Fuelled by a neural-network-based model, our interface can decode finger-wise forces in real-time with 3.3\% mean error, and generalize to new users with little calibration. Through an interactive psychophysical study, we show that human perception of virtual objects' physical properties, such as stiffness, can be significantly enhanced by our interface. We further demonstrate that our interface enables ubiquitous control via finger tapping. Ultimately, we envision our findings to push forward research towards more realistic physicality in future VR/AR.
\end{abstract}

\ccsdesc[500]{Computing methodologies~Virtual reality}
\ccsdesc[500]{Computing methodologies~Mixed / augmented reality}
\ccsdesc[500]{Computing methodologies~Perception}
\ccsdesc[500]{Computing methodologies~Neural networks}
\ccsdesc[500]{Human-centered computing~Haptic devices}

\keywords{Electromyography, Force-Aware Neural Interface, Machine Learning, Haptic Perception}

\begin{document}
\maketitle
\section{Introduction}
\label{sec:introduction}

The \emph{visual} gaps between real world and virtual environments have been rapidly shrinking with the advance of novel display and rendering technologies. However, developing matching realistic-\emph{feeling} interfaces that let users interact as if they were in the physical world, stands out as a chronically persistent and doggedly resistant challenge \cite{10.1145/3486184.3491079}. Physical interactions, such as lifting, grasping, brushing, pushing, and prodding involve a bi-directional interchange between humans and the environment: our muscles exert forces on objects, while we perceive the visual (and sometimes haptic) feedback response in the objects' reactions. To establish the same loop in virtual environments, researchers have devoted extensive effort to advance the quality of feedback sensations with haptic devices and rendering methods. However, it has long remained difficult to transfer real-world physical human applied forces of dexterity and agility into convincing virtual form. This incomplete loop leaves VR/AR disadvantaged in its ability to faithfully and convincingly represent real experiences.

The idea of directly sensing, tracking, and decoding user-induced forces has emerged as a promising line of research, with the implication that this information provides a scaffold for building natural and intuitive interaction experiences \cite{ernst2002humans,bergstrom2019human}. Wearable force sensors now provide high-precision and high-resolution data \cite{sundaram2019learning,luo2021learning}. However, existing force-sensing technologies are often bulky, wired, and directly attached to hands. This hampers their applications for natural interaction and makes the devices undesirable as consumer-level interfaces. One solution has been to skip devices altogether. Purely data-driven visual-to-force learning methods \cite{Ehsani_2020_CVPR} have been proposed, allowing for contactless \emph{estimation} of user force. However, they may suffer from occlusions, low precision, and action-perception delays due to the high load of transmission and processing.

The advancement of neural sensing enabled central (from the brain \cite{anumanchipalli2019speech,willett2021high}) and peripheral (from the muscles \cite{salemi2021rotowrist,liu2021neuropose}) solutions for decoding human action intentions, as electrophysiological responses. However, decoding the intended force for interaction has been so far unsolved due to the variance across human users, the lack of correlated data, computational complexity that blocks real-time performance, and the inevitable pervasive sensory noises affecting biological signals \cite{hof1991errors}.

We introduce an end-to-end neural interface that reduces the physicality gap between real experience and VR/AR. The result is a real-time system for \emph{dexterity-enabled force-aware VR/AR}. The system's chief advantages are that, (1) it allows for natural, unimpeded, forearm and hand movement; (2) using off-the-shelf electromyography sensors; (3) with low latency for force-and-response interactions with computer graphics; (4) in ways that are generalizable across a diversity of users.

Our research shows that very detailed and rich physical experiences of manual dexterity can be delivered to VR/AR systems and paired with high-fidelity graphics for visual similitude, in ways that neatly and realistically close the loop between intention and interaction in VR/AR experiences. Our system offers a tractable solution to existing bottlenecks in directly sensing and resolving users' physical intentions in VR/AR systems, using machine-learning on skin-surface electromyography (sEMG) sensors to identify, track, and decode signals of physical activity at rates that allow for matching design and delivery of experiential content in VR/AR. These developments, while preliminary, open-up new pathways for VR/AR experience in gaming, design, and object control. While we describe the research, development, and evaluation pipeline, we note that the system is application-ready. We demonstrate practical examples on low-cost commercially available sEMG sensors and widely used VR/AR technology. We will also open-source both our dataset and source code to the community to support future work.

While the system is shown to work parsimoniously in user-testing and evaluation, the research behind it is non-trivial. To develop the proposed neural interface, we start by collecting a large-scale joint dataset via force-sensing and surface EMG devices. The dataset consists of the time-synchronized signals between fingertip forces and the corresponding EMG signals. By leveraging our specialized dataset, we developed the first real-time learning-based framework that tracks and decodes human physical forces from multichannel muscle activation signals. The dataset is populated, initially, using a set of participant experiments to record EMG signals from forearm muscles while participants directly perform various natural hand-object interactions, such as pressing and pinching. On this initial dataset, we trained a convolutional neural network (CNN) model on the frequency-transformed signals to robustly learn the complex mapping between muscle activities and actions. The trained model isolates the force-induced bio-electrical signals from hand motions and estimates the forces exerted at the fingertips. During run-time, the model only uses the \emph{past} 624ms of EMG data, enabling low-latency force inference in real-time. With this model on hand, we show that only minimal calibration is required to transfer and generalize it to unseen users.

In order to validate the system we have conducted a systematic user study and evaluated users' experience during interaction with virtual objects. We will present a series of psychophysical experiments and objective analysis that reveal our system to be robust and generalizable. Moreover, we will show that the system can enhance users' perceptual understanding of virtual objects' physical and material characteristics in VR, by extending their capabilities for natural human interaction with graphical objects. Our experimentation also demonstrates that the system is broadly resilient to variation in user physiology, sensor placement, and tasks.

In summary, this paper contributes:
\begin{itemize}
    \item An end-to-end EMG-based neural interface that decodes, transfers, and applies hand-induced forces with low-latency in VR environments;
    \item A prototype interaction system that leverages our method to enhance human's perceptual understanding of material characteristics in VR;
    \item A real-time and generalizable CNN-based model established in the frequency demain of EMG-sensed muscular potentials with a force-tailored loss design;
    \item A set of user experiments to demonstrate the generalizability of the system to perturbations in sensor placement, shifting task context, and uniqueness of users;
    \item Proofs of concept for natural interaction with computer graphics in VR.
\end{itemize}

We provide the source code for our force regression models, real-time interaction system, and accompanying EMG-Force dataset at \url{https://github.com/NYU-ICL/xr-emg-force-interface}.
\section{Related Work}
\label{sec:prior}

\subsection{Biometric Sensing for Immersive Interaction}

Accurately sensing human behaviors is fundamental to favorable human-environment interaction. With recent advancements of various sensing technologies, both in hardware and software, we are now entering an era where unprecedented means of multi-modal interaction with virtual environments are possible. For instance, eye tracking enables real-time foveated rendering~\cite{patney2016towards,kim2019foveated} and enhances VR redirected walking~\cite{sun2018towards,langbehn2018blink}; face tracking generates lifelike virtual avatars for telecommunication~\cite{ma2021pixel,chu2020expressive,chen2021high}; whole-body tracking allows for intuitive control and feedback for virtual interaction~\cite{newell2016stacked,cao2017realtime,joo2018total,kanazawa2018end}. In virtual environments, users largely rely on hand-based interfaces for interaction, making hand behaviors particularly indicative of their intention and status. As a result, hand tracking has attracted considerable research interest in computer graphics~\cite{romero2017embodied,wan2018dense,boukhayma20193d,han2020megatrack}. However, tracked position information alone is insufficient to achieve immersive VR experience. Without the \emph{feeling} of hands, positional tracking essentially casts ghost appendages in users' field of view. This misses the sense of \emph{corporeality} and thus the sense of capabilities that humans feel as they use their hands in the real world. We reason that \emph{hand-induced interaction force} is another indispensable component of human embodiment in virtual scenes that is often overlooked or only approximated~\cite{pham2015towards,zhu2016inferring} in prior works.


\subsection{Sensing and Interacting with Contact Forces}

Contact forces are an essential modality for understanding and enhancing human-object interaction~\cite{sundaram2019learning,luo2021knitui}. Unlike visual stimuli, force information must be communicated in a \emph{two-way} fashion when we interact with and establish understanding of virtual environments. While users apply forces to a virtual object, they also receive haptic feedback from the object's response~\cite{gonzalez2021x,dangxiao2019haptic,yoshida2020pocopo}. For the latter, which has been addressed in computer graphics as haptic rendering~\cite{lin2008haptic}, researchers have explored various ways of applying tactile effects to users, ranging from grasping and touching~\cite{choi2018claw,verschoor2020tactile,choi2016wolverine} to texture~\cite{benko2016normaltouch}, shear~\cite{whitmire2018haptic} and gravity~\cite{choi2017grabity}.

However, the inverse problem of naturally sensing and exploiting human-exerted forces in the context of VR remains an open challenge. Existing solutions are either based on hand-held input devices or force-sensing wearables~\cite{sundaram2019learning,luo2021learning}. While such methods can provide high-precision force measurements during hand-object interaction, their \emph{obtrusive} design inevitably compromises finger dexterity, increases the \emph{frictions} between users and virtual environments, and limits their availability for daily usage. To develop a natural and intuitive force interface for VR, we attempt to sense hand-applied forces from the controlling muscles located on the forearm by leveraging the biological mechanism of human hands as described in~\Cref{sec:method-anatomy}. This allows us to completely bypass on-hand measurements and achieve force-enabled VR interaction in a natural bare-hand manner.


\subsection{EMG-Based Human-Computer Interface}

Recent advancements in neural interfaces have demonstrated the great potential of interactive devices that directly interface with the human body and interpret neuronal activities for downstream tasks~\cite{anumanchipalli2019speech,willett2021high,hochberg2012reach,flesher2021brain}. Among these interfaces, EMG has emerged as a promising interaction medium, especially in VR and AR~\cite{tsuboi2017proposal,hirota2018gesture,koniaris2016iridium}. A major benefit of EMG for immersive interaction is the potential that it offers for completely bypassing the often-used solution of camera-based tracking, which has serious side effects of being open to limitation by occlusions and field of view~\cite{pai2019assessing}. To advance EMG approaches, considerable research efforts have been made to infer hand poses from forearm EMG, including gesture recognition~\cite{du2017semi,rahimian2021few,gulati2021toward,sun2022deep,javaid2021classification,jo2020real}, hand orientation estimation~\cite{andrean2019controlling,zhao2020emg}, and finger tracking~\cite{liu2021neuropose,qi2021active,zhang2022simultaneous}. The knowledge may then be leveraged towards camera-free VR control~\cite{ahsan2009emg}. We reason, additionally, that tracking hand-object \emph{interaction forces} is indispensable to creating realistic physical effects in VR, e.g., via physics-based simulation methods. However, finger-exerted forces are continuous, transient, subtle, and changeable, thus pitching fundamental challenges for decoding.

Prior research investigated the possibility of estimating hand/finger-level forces from forearm EMG~\cite{liu2013emg,castellini2009surface,castellini2012using,gailey2017proof,zhang2022simultaneous,fang2019attribute,baldacchino2018simultaneous,mao2021simultaneous,bardizbanian2020efficiently,bardizbanian2020estimating,cho2022training,becker2018touchsense,wu2020semg,hu2022novel,wu2021optimal,martinez2020grasp,martinez2020online}. Despite exciting preliminary results, deploying them in practical VR applications is still in its infancy. Several open problems remain mostly unresolved:

\paragraph{Flexibility for real-life usage}
Existing solutions commonly assume controlled laboratory settings. For example, work presented by Castellini and Koiva \cite{castellini2012using} can only operate when the user's hand is artificially pinned and constrained in a flat wooden mold/guide. In the approach by Zhang et al. \cite{zhang2022simultaneous}, it is necessary to place hard-wired electrodes up and down an entire arm; moreover, force detection relies on an elaborate mechanical metal force-sensing device that is hard-bolted to a table. In the approach shown by Baldacchino et al. \cite{baldacchino2018simultaneous}, there is also a requirement that an entire arm be fitted with electrodes. These systems are fantastic early proofs-of-concept, but interacting in VR demands requires that free-form interaction is supported---we would argue that it also needs to be as natural as possible---and this necessitates a different approach over the current state-of-the-art to successfully bypass those complications. Our approach introduces a relaxed, accessible, natural test-bed that can accommodate freely realistic postures and gestures of the hand. The level of authenticity that we have achieved relative to real-world hand and finger forces contrasts with much of the prior art. Existing approaches are highly isometric, which artificially limits free interaction in testing and in use.

\paragraph{Simultaneous and continuous multi-finger force measurement}
To reproduce natural dexterity, it is necessary to enable all fingers to operate together and apply varied forces simultaneously. This is critical to how we use our upper limbs to manipulate and explore the world around us. Most existing approaches to reproducing this in VR have focused on generalized hand-scale gestures~\cite{hu2022novel,fang2019attribute,gailey2017proof,wu2021optimal,martinez2020online}. Humans rely on most muscles in the forearm and neural control of these muscles produces electrical signals. These signals are notably unambiguous and thus open to direct detection. This is not always straightforward outside of clinical sensing. Beyond a classification problem, regressing the exact force value positions presents additional challenges due signal noise and individual variances. This has been tried before. For example, Baldacchino et al. \cite{baldacchino2018simultaneous} presented a regression approach, but rather than sensing they tackled the challenge through data science on an existing database \cite{atzori2015ninapro}, which was limited to nine variations of a simple (and \emph{single}) finger-on-surface pressing motion (compare this to the free-form and multi-finger dexterous gestures that our scheme tackles). Real humans of course use their fingers as they please during dexterous tasks; limiting dexterity to a single finger would seriously hamper usability. Continuity in the temporal domain presents another challenge. As one can imagine, the signal firing of muscles in the forearm is highly dynamic. Previous work have addressed this by approximating dynamics as ``action shifts" between hand postures \cite{gailey2017proof}. This is really just a workaround that substitutes state transition for actual dynamics. This is problematic for VR settings, which are often highly dynamic, with users that are usually quite aware of how fast their hands and fingers move in the real world. Building realistic and fast-adaptive temporal continuity between user actions and responding force-aware graphics is therefore critical in supporting natural interaction.

Our approach, by comparison, \emph{simultaneously} isolates and teases out signal details for \emph{individual fingers} with \emph{spatio-temporally continuous} force prediction. Our aim, in doing so, is to support a wide range of natural interactions in VR/AR, including those that require fine-grain dexterity maneuvers. The force-based dexterous abilities that are accessible via our scheme (e.g., finger tapping) are well beyond the capabilities of existing prior art (which focus on finger posture (not force) or track very stylized dexterity such as simple pressing actions). This is achieved via our machine learning approach and an in-house dataset with robust interaction variety.

\paragraph{Generalizability} 
Daily interactive scenarios require generalizable systems for everyone, without tedious pre-usage preparation. The prior art in this domain adopts an approach that validates cross-user force prediction accuracy with datasets under \emph{identical settings}~\cite{castellini2009surface,zhang2022simultaneous,bardizbanian2020estimating,mao2021simultaneous,bardizbanian2020efficiently,becker2018touchsense}. By contrast, our \emph{frequency domain neural network method} tackles the long-looming generalizability problem in EMG data decoding. In this paper, we show that we can collect data on two completely different days (with associated shifts in placement of sensors) as well as for completely different users (with shifting reactions, varying arm and finger morphology, and different dexterity and skills) with less than 2 minute calibration. Solving sensing generalizability and subject generalizability---in tandem---is a significant contribution to the literature. Moreover, it greatly expands the applicability of our scheme for VR/AR, where there will necessarily be wide variation gaps in sensing conditions and users.

\section{Method}
\label{sec:method}

In the following, we first review the biological mechanism of human hands and illustrate how muscles on the forearm control finger-level forces in~\Cref{sec:method-anatomy}. Then, we describe how we tailored our EMG-force joint data collection to capture the complex mapping from muscle activations to finger-level forces in~\Cref{sec:method-data-collection}. Finally, using the established dataset, we detail our frequency-domain muscular force learning pipeline, with joint classification and regression, in~\Cref{sec:method-model}.

\begin{figure}[t]
\centering
\includegraphics[width=0.99\linewidth]{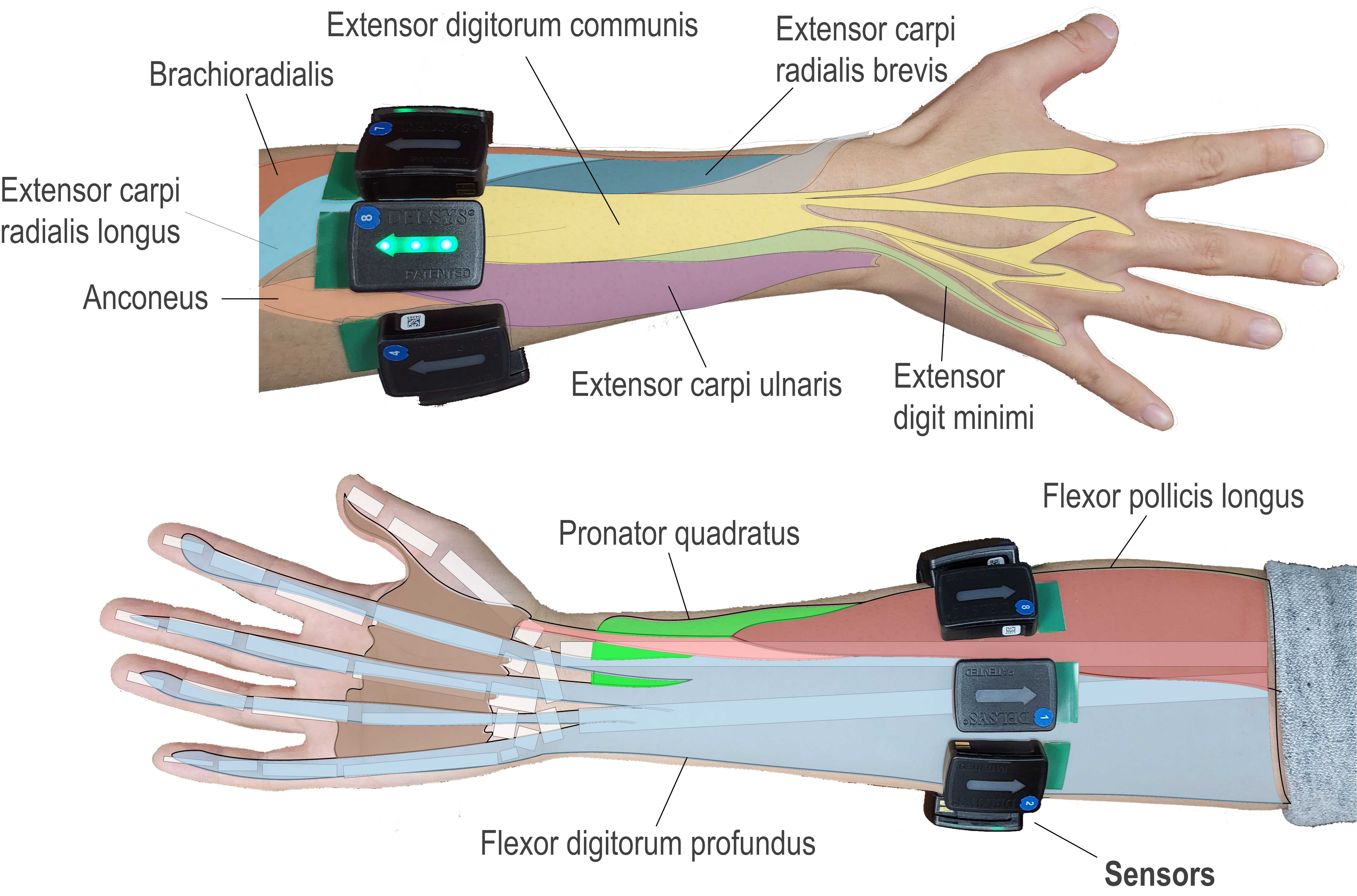}
\Caption{Anatomical illustration of forearm muscles controlling the flexion and extension of thumb and four fingers.}
{Our system predicts hand-induced forces at finger level by sensing forearm muscle activations with EMG sensors, preserving the dexterity necessary for delicate hand activities in VR/AR.}
\label{fig:method-anatomy}
\end{figure}


\subsection{Biological Model of Human Hands}
\label{sec:method-anatomy}

To bypass the limitations of passively measuring hand-exerted interaction forces using cumbersome (and interfering) sensors such as gloves, we argue that such information can be actively decoded at the finger level from the bioelectric signals reflecting forearm muscle activations, which wireless EMG sensors can in turn capture.

\paragraph{Hand-forearm joint biomechanical mechanism}
As shown in~\Cref{fig:method-anatomy} (bottom), hands, the most dexterous limbs on the human body, exhibit high degree-of-freedom (DOF) articulations through a large number of finger joints, allowing us to perform complex and subtle interactions with the surroundings. The muscles driving this delicate articulated structure are: 1) extrinsic muscles spread over the anterior and posterior compartments of the forearm; 2) intrinsic muscles located right in the hand.

From the perspective of VR/AR applications, users most often interact with their surroundings through bare-hand touching, pressing, pinching, and gripping~\cite{von2001bare}. The major contributing muscles in these interactions include flexor digitorum superficialis, flexor digitorum profundus, and flexor pollicis longus, all residing in the anterior compartment of the forearm. In particular, flexor digitorum superficialis controls the flexion of PIP and MCP joints for the 4 fingers and the wrist; flexor digitorum profundus controls the flexion of DIP joints for the 4 fingers as well as the flexion of MCP joints for the 4 fingers and the wrist; flexor pollicis longus controls the flexion of IP and MCP joints for the thumb.~\Cref{fig:method-anatomy} illustrates the anatomical structure of these forearm flexor muscles. By investigating the signals passed when invoking these biomechanics, we reason that it becomes possible to \emph{sense and learn} hand operations from the connected forearm.

\paragraph{Bioelectric mechanism}
Muscles are composed of constituent elements called motor units, and the contraction of each single muscle is managed by a specific group of motor units. On the other hand, motor units are made from more fundamental units called muscle fibers. When activated by our brain, muscle fibers within the same motor unit fire together and generate a propagating electrical potential called motor unit action potential (MUAP) via the elevation of $Ca^{2+}$ in the sarcoplasm \cite{melzer1984time}. 
When placed on our skin, the EMG sensors record such electrical signals in real-time. The various hand-object interactions that we can perform are results of varying activation patterns of the involved muscle fibers, which are themselves reflected by the recorded signals. However, how analytically or numerically the electrical signals are coupled with mechanical forces remains an open challenge, especially given the inevitable sensing noise. In the following section, we discuss our attempts toward a robust electric-mechanic signal decoding in the frequency domain.


\subsection{EMG-Force Joint Data Collection}
\label{sec:method-data-collection}

\begin{figure}[t]
\centering
\includegraphics[width=0.9\linewidth]{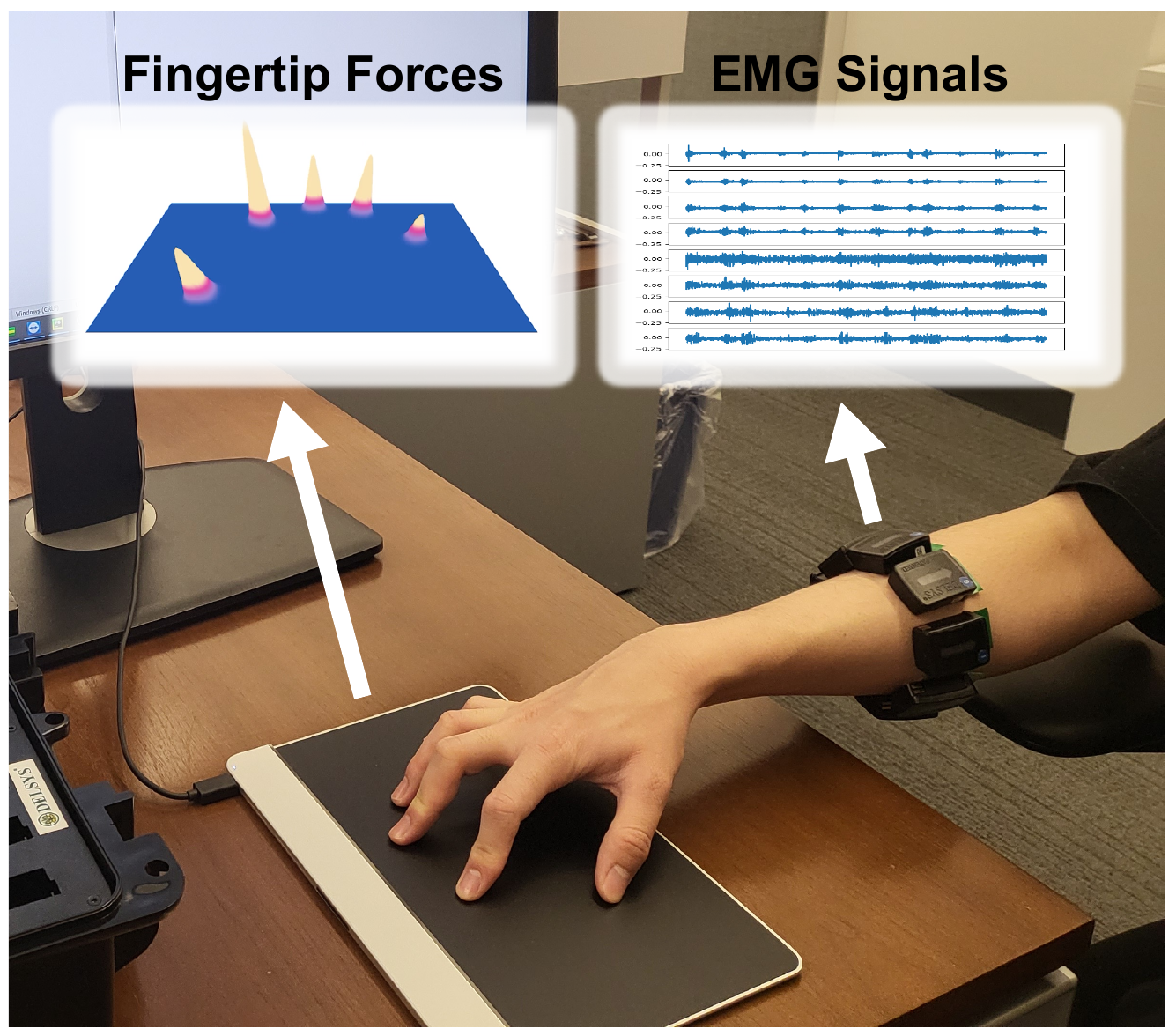}
\Caption{Experimental setup for collecting time-synchronized EMG and force data.}
{During the data collection, participants were asked to interact with a Morph Sensel trackpad through various pressing and pinching actions while wearing 8 EMG sensors on his/her right forearm.}
\label{fig:method-data-collection}
\end{figure}

We aim to establish a bioelectrical-mechanical bridge via a data-driven approach. To this end, we first collect time-synchronized EMG and force data in a supervised manner. To collect EMG electrical signals, we adopt $8$ Delsys Trigno EMG sensors (from Delsys Inc, USA) and overlay them on the forearm in a way that all muscles of interest are monitored. All $8$ EMG sensors are wirelessly synchronized at 2000 Hz. We note that this is a factor of \emph{ten times} the bandwidth of the (now discontinued) Myo sensor that is used in prior art, e.g., Javaid et al. \cite{javaid2021classification}. In their review of the accuracy of sEMG sensors, Pizzolato et al. \cite{RN10381} discuss this issue directly, noting that ``the Myo is not suited to record high quality sEMG signal data including the full power spectrum of sEMG (that can include frequencies of up to 300-500 Hz)'' (p.10). Our captured data are streamed to a desktop computer over WiFi in real-time. To collect finger-wise force data, we employed a Morph Sensel trackpad with pressure sensors arranged into a dense array. The data collection setup is illustrated in~\Cref{fig:method-data-collection}. In particular, we divide the tracking area into 5 non-overlapping regions so that each fingertip only taps onto its dedicated partition throughout the data collection process. Detected contact points with force information can then be correctly attributed to the corresponding fingers. Note that this design choice is only adopted to ease force labeling efforts and we do not assume any specific wrist/finger poses during either training or testing. Also, the dividing strategy is user- and motion-specific to accommodate the hand size and personal habit of different users. The data collection code for both modalities is launched using multi-threading, and the system timestamps are exploited for overall synchronization.

\begin{figure*}[t]
\centering
\includegraphics[width=0.99\linewidth]{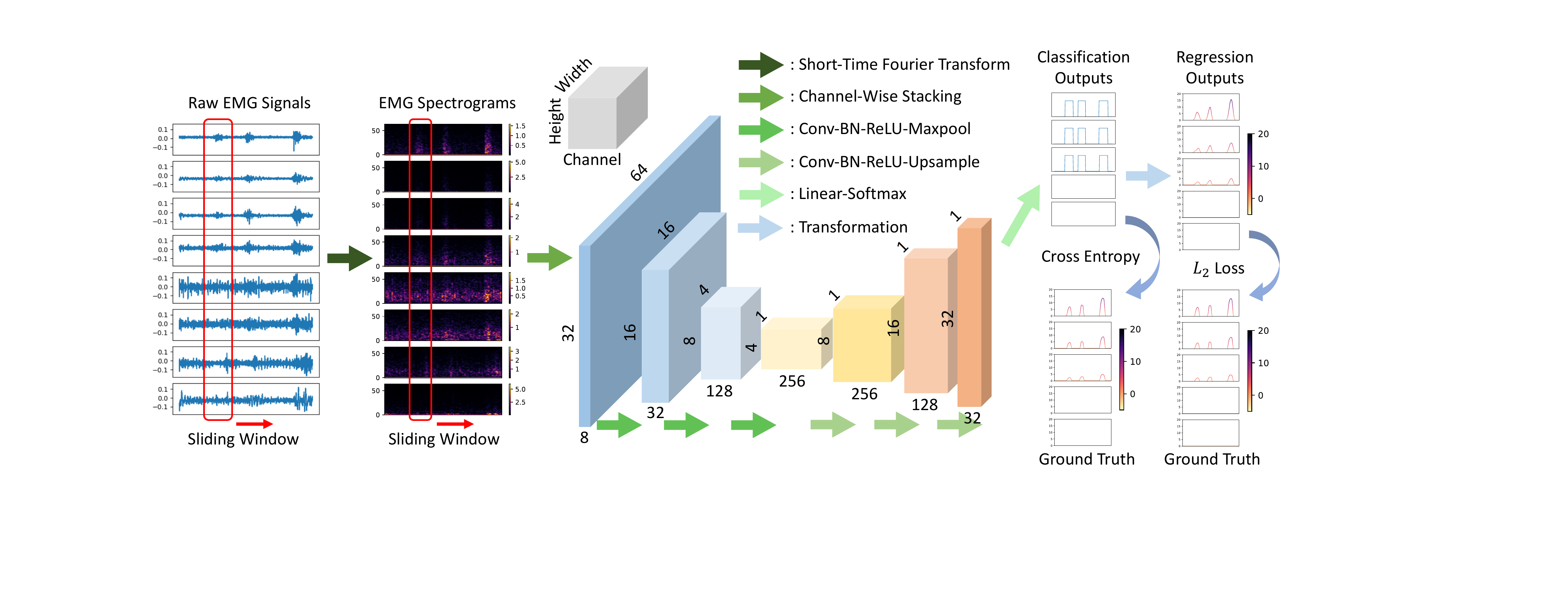}
\Caption{Illustration of the deep learning pipeline embedded in our system.}
{To optimize for parameter efficiency and resilience to data noise, we transform raw EMG signals into frequency domain via STFT, treat the resulting spectrograms as multi-channel images, and employ a lightweight CNN model with bottleneck design. Training is performed using a customized classification-regression joint loss tailored to the task of force estimation. When put into action, our system uses a fixed-size sliding window to retrieve the latest frames from wirelessly streamed EMG signals and decodes finger-wise forces in real time.}
\label{fig:method-model}
\end{figure*}

\paragraph{Transformed data representation}
As a typical bio-electric sensor, EMGs also suffer from a certain level of measurement noise including powerline noise and other electromagnetic artifacts. Existing EMG processing approaches typically extract and learn from statistical features in the time domain, such as mean absolute value, average amplitude change, interquartile range, etc \cite{spiewak2018comprehensive}. Consequently, subtle noise or distortion may cause significant feature-space error \cite{boostani2003evaluation}, harming the change-sensitive force-bioelectricity correlation.

Drawing inspirations from audio research, we compute the spectrograms of EMG signals using short-time Fourier transform (STFT) so that high-frequency additive noise may be more distinctly isolated. Another computational advantage of learning with frequency-domain representation is that the EMG signal from each electrode, or channel, is now a 2D array instead of a 1D time series and that we can seamlessly take advantage of powerful convolutional neural network (CNN) models for better parameter efficiency and generalization capability. Specifically, we adopt a Hanning window of size 256 sample points, which corresponds to a duration of 128ms, with hop length set to 32, to obtain 129 frequency bins. In addition, a resampling step is needed to temporally align raw force data (the sampling frequency of Morph Sensel is around 125Hz) with computed EMG spectrograms. A nearest-neighbor-based interpolation is adopted for this purpose.


\subsection{Muscular Force Learning Pipeline}
\label{sec:method-model}

A main roadblock for EMG sensors is the well-known challenge of aligning the electrodes exactly on muscles. For instance, as seen in~\Cref{fig:method-anatomy}, sensors may commonly cross-ride on or fall in the gap between the underlying interwoven muscle bundles. As a result, although the activation information of all target muscles are captured by EMG sensors, directly assigning the electric signals to individual muscle-group and joints becomes unrealistic. To robustly recover finger-wise force information from raw EMG data, we resort to the data-driven paradigm and adopt powerful neural network models to learn this highly non-linear correlation between forearm EMG signal and finger-wise forces.

\begin{figure*}
\centering
  \includegraphics[width=0.99\linewidth]{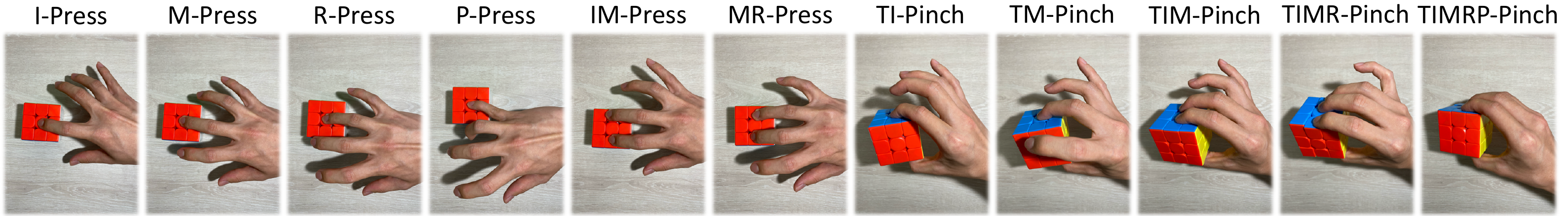}
\Caption{Types of hand-object interaction selected for constructing our EMG-Force dataset.}
{Capital letters before the hyphen, namely T, I, M, R, and P, stand for thumb, index finger, middle finger, ring finger, and pinky finger, respectively.}
\label{fig:evaluation-dataset}
\end{figure*}

\paragraph{Model architecture}
While recurrent neural network (RNN) has been a common practice for sequential data learning, recent advancements in audio learning have shown that deep CNN models with properly processed input data are capable of delivering better performance in some cases, thanks to their highly efficient parameter usage which allows for very deep design \cite{oord2016wavenet}. We are inspired to exploit convolutional filters to extract deep features from the 2D spectrograms. While our input data points live in a high-dimensional space ($129$), those features containing the semantic information of finger-wise forces are embedded in a subspace of much lower dimension. 
To efficiently extract relevant information and mitigate overfitting to training data, we employ an encoder-decoder architecture to enforce a low-dimensional latent space. Also, we only feed the $64$ low-frequency components from the spectrograms to the encoder model to remove high-frequency data noise and allow for accelerated performance. In addition, we provide the model with sequential data from a long time interval (32 consecutive frames in the spectrograms, which correspond to 624ms raw EMG data) instead of a single frame to let it exploit information from previous frames and better satisfy temporal constraints. The input data size is thus $(N, C, L, S)$, where $N$, $C=8$, $L=32$, and $S=64$ denote the batch size, number of sEMG channels, input sequence length, and number of EMG frequency components, respectively. The encoder model consists of repeating Convolution-BatchNorm-ReLU blocks, with each block followed by a $2\times2$ Maxpool layer for downsampling in time and frequency dimensions. Similarly, the decoder model also consists of repeating Convolution-BatchNorm-ReLU blocks, with each block followed by a $2\times2$ bilinear Upsample layer for increasing time dimension. After that, the decoder output is transformed by a linear layer in the channel dimension to match desired force outputs, e.g. 5 values for 5 finger-wise forces. The output data size is thus $(N, F, L, 1)$, or $(N, F, L)$ with the fake frequency dimension squeezed out, where $F$ denotes the pre-defined number of force components. All convolutional layers have kernel size $3\times3$. When putting the model in action, spectrograms of streamed EMG signals are computed on the fly and a sliding window of length $L=32$ feeds the latest data to the model for real-time inference. Detailed model architecture, data flow and input/output dimension at each layer are illustrated in~\Cref{fig:method-model}.

\paragraph{Joint classification and regression}
Estimating continuous finger-wise forces, by its nature, is a regression problem, and it is natural to adopt common regression losses, such as $L_{1}$ or $L_{2}$ loss, as the objective function. However, the muscle-generated forces in interactive scenarios have two unique patterns: humans apply forces only sparsely in the real-world; and the variance of force levels is commonly high, ranging from light touches to hard pushes. In practice, regression-based learning oftentimes tends to predict non-zero values (false positive when we do not generate forces) or over-smooth low-amplitude values (false negative for light forces). For our targeted VR/AR applications, this seemingly small estimation error can lead to visually noticeable artifacts and largely compromise user experience (e.g., causing constant vibrations on objects or producing no reaction on low-force touches).

A na\"ive solution to this problem is to set a cut-off threshold such that the estimated values below it are treated as zero. Although this modification enables zero-value output, tweaking the threshold can be unworkable in practice and the performance is still barely satisfactory as will be shown in~\Cref{sec:evaluation-learning}. To address this issue, we introduce a classification loss to better differentiate between EMG sequences with and without forces. Specifically, for each time frame $t$ and each finger $i$, the model outputs a value $p_{i}^{t} \in [0, 1]$ indicating the probability of that finger applying force at that time frame. A cross entropy loss $L_{c}$ is employed to train the model for this force/no-force binary classification task. On top of $p_{i}^{t}$, we compute the predicted force as $\hat{F}_{i}^{t} = 2F_{\max} \cdot \max(0,p_{i}^{t}-0.5)$, where $F_{\max}$ denotes the force upper-bound and defines the predicted force range. A $L_{2}$ loss is then employed to train the model for force regression.
\begin{equation}
\label{eq:classification-loss}
    L_{c} = \frac{1}{T \cdot I} \sum_{t=1}^{T} \sum_{i=1}^{I} y_{i}^{t} \cdot \log p_{i}^{t} + (1-y_{i}^{t}) \cdot \log (1-p_{i}^{t})
\end{equation}
\begin{equation}
\label{eq:regression-loss}
    L_{r} = \frac{1}{T \cdot I} \sum_{t=1}^{T} \sum_{i=1}^{I} \|\hat{F}_{i}^{t} - F_{i}^{t}\|^{2}
\end{equation}
where $y_{i}^{t}$ and $F_{i}^{t}$ denote the ground-truth force label and value, respectively, for finger $i$ at time frame $t$.

A hyper-parameter $\lambda$ is introduced to balance between classification and regression, and the overall loss $L$ takes the form:
\begin{equation}
\label{eq:total-loss}
    L = L_{c} + \lambda \cdot L_{r}
\end{equation}
The joint loss above is designed such that, for finger $i$ at time frame $t$: when $p_{i}^{t} \in [0,0.5]$, we have $\hat{F}_{i}^{t}=0$, only $L_{c}$ takes effect and the model focuses on correcting wrong classifications; when $p_{i}^{t} \in (0.5,1.0]$, we have $\hat{F}_{i}^{t} = 2F_{\max} \cdot (p_{i}^{t}-0.5) \in (0,F_{\max}]$, $L_{c}$ and $L_{r}$ together push the model towards the joint classification-regression goal. As a result, we are able to not only get zero-value outputs when there is no force, but also prioritize classification over regression at the beginning stage of training, since the estimated force value will be useless if the predicted class is wrong in the first place. Note that $p_{i}^{t}$ is exploited to both differentiate between no-force and force (classification with threshold $p_{i}^{t}=0.5$) and compute the predicted forces $\hat{F}_{i}^{t}$ (regression).
\section{Evaluation}
\label{sec:evaluation}

To evaluate our method and system, we first discuss in~\Cref{sec:evaluation-dataset-metrics} the specifics of our EMG-Force dataset and the evaluation metrics for quantifying the performance of our CNN-based regression model (detailed in~\Cref{sec:method-model}). Then, we present the results of fingertip force estimation for various common hand-object interactions in~\Cref{sec:evaluation-learning}. Following that, we compare our approach with existing vision-based methods in~\Cref{sec:evaluation-comparison}. We further study the time-efficient generalization of the pre-trained model to new users in~\Cref{sec:evaluation-generalization}. In addition, we also analyze the resulting neural interface in terms of latency and storage for real-time applications in~\Cref{sec:evaluation-system}. Finally, we conduct a user study to demonstrate the knowledge of contact force value could benefit material perception and enhances physical realism for real-world VR/AR interaction in~\Cref{sec:evaluation-user-study}.


\begin{figure*}
\centering
  \subfloat[Action-wise and overall performance of user-independent model.]{
    \label{fig:evaluation-learning-action}
    \includegraphics[width=0.99\linewidth]{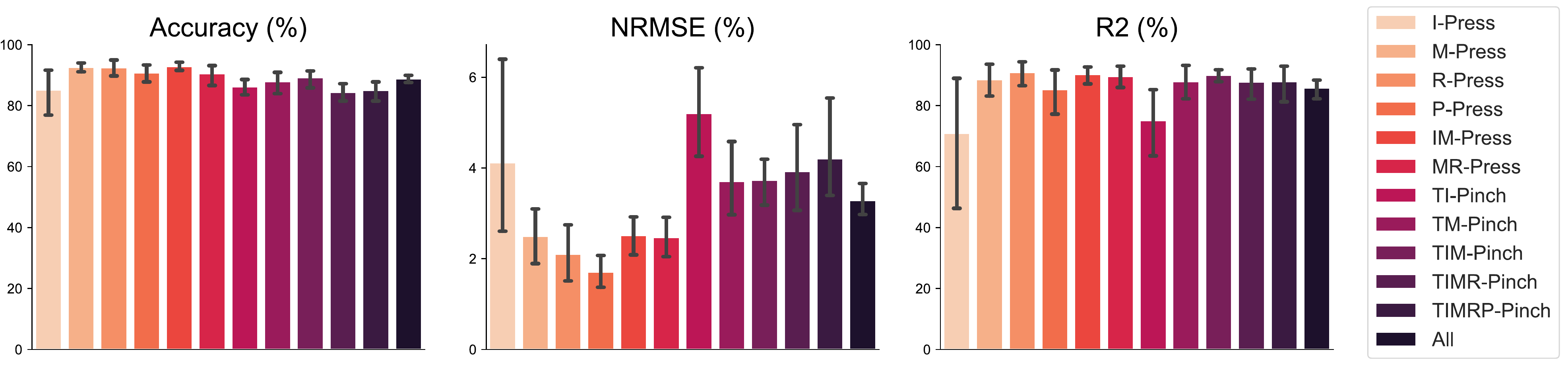}
  } \\
  \subfloat[Subject-wise and overall performance of user-independent model.]{
    \label{fig:evaluation-learning-subject}
    \includegraphics[width=0.99\linewidth]{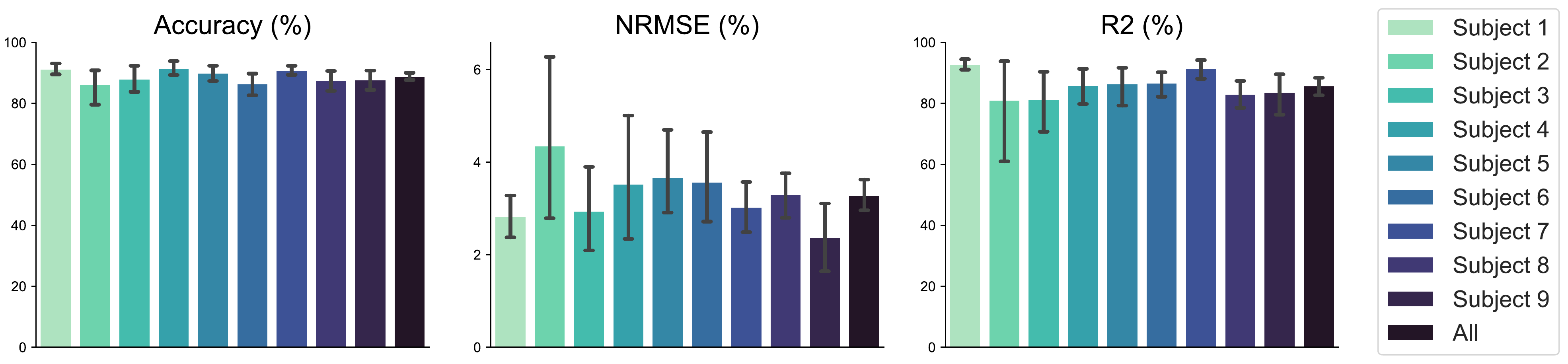}
  }
\Caption{Objective evaluation of user-independent model.}
{\subref{fig:evaluation-learning-action} shows the action-wise performance of user-independent model in estimating finger-wise forces, with $95\%$ confidence intervals overlaid. Similarly,~\subref{fig:evaluation-learning-subject} shows the subject-wise performance.}
\label{fig:evaluation-learning}
\end{figure*}

\subsection{EMG-Force Dataset and Evaluation Metrics}
\label{sec:evaluation-dataset-metrics}
 
The relationship between forearm muscle activations and finger actions exhibits a highly complex mapping~\cite{farina2016characterization}. On top of this complexity, its variations across subject identity, arm/hand posture, and subject's physical condition further add to the complexity of its precise characterization. Additionally, the electric signal detected by each EMG sensor is inevitably a superposition of multiple muscles' activities (as shown in~\Cref{fig:method-anatomy}), which only makes decoding finger-wise forces from EMG signals even more difficult. Therefore, it is crucial to establish a comprehensive training dataset covering common and natural hand-object contact patterns, so that the neural network model can effectively capture this relationship and acquire better generalization capability. Most prior art relies on the NinaPro dataset \cite{atzori2015ninapro} which is actually intended for manipulating robotic arms and is collated from the CyberGlove data glove (and therefore not representative of natural hand or finger movements). Here, we introduce an alternative data set that we have collected ourselves.

\paragraph{EMG-Force dataset}
When users perform hand-object interactions, whether in the physical or virtual world, pressing, pushing, pinching, and holding are arguably among the most frequent actions~\cite{ingram2008statistics}. These actions allow users to not only better perceive surrounding objects, especially their physical properties, but also pick them up for further interactions. Based on their respective force exertion mechanism, we partitioned these actions into two representative groups: pressing/pushing and pinching/holding. Eleven common finger combinations were selected for data collection purposes, with six for the former and five for the latter. This set of actions, which we call action set $\mathcal{A}$, is summarized in~\Cref{fig:evaluation-dataset}. To build our EMG-Force dataset, we recruited $9$ participants (ages $21-30$, $5$ females, $4$ males). Following the data collection and pre-processing pipeline described in~\Cref{sec:method-data-collection}, we conducted three collection sessions with each subject, capturing $30$ seconds of data for each action during each session. In total, each participant contributed $990$-seconds of time-synchronized EMG and force data. For each subject, $2$ randomly selected sessions (out of $3$) were earmarked for the construction of the training set. The remaining session was withheld and only used for evaluation. When performing pinching actions during data collection, participants were instructed to keep their four fingers over the trackpad and their thumbs below the table, so that they could pinch the ensemble of trackpad and table in a natural manner. Besides, they kept the resultant force stable and balanced (i.e., $\sim0$N). The ground-truth forces for the four fingers were directly recorded by the trackpad, and the force for the thumb was derived as the additive inverse. With the importance of data coverage in mind, all subjects were instructed to randomize their force intensity level within the natural range of each action. In addition, a random spacing in time was enforced between adjacent interactions, so that neural network models do not over-fit to unintended temporal features. The EMG-Force dataset contains light touch less than 1N and firm press up to 30N, covering the typical functional force range of human fingers~\cite{xu2020real}. In particular, the maximum force for the five fingers in Newton, from the thumb to the pinky finger, are $29.8$, $24.4$, $25.6$, $20.4$, and $14.7$. The mean/standard deviation/interquartile range are $10.1/8.1/13.9$, $5.8/4.6/7.4$, $5.7/4.7/7.6$, $3.4/3.1/4.0$, and $2.6/2.5/3.2$. Our CNN model predicts force values in $[0, 30]$, which is configured through the force upperbound \(F_{\max}\).

\begin{figure*}
\centering
  \includegraphics[width=0.99\linewidth]{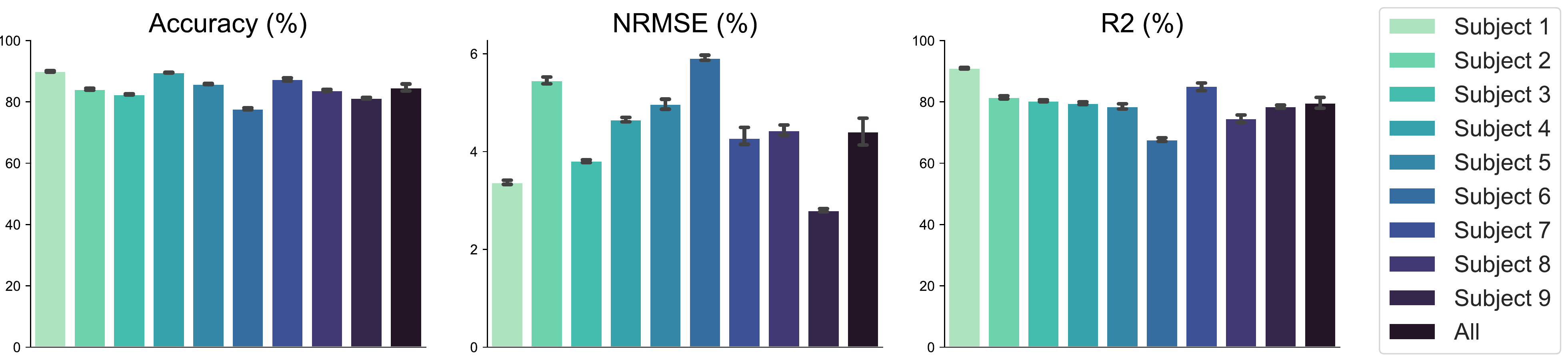}
\Caption{Objective evaluation of user-specific model.}
{Each user-specific model is calibrated using 165-second synchronized EMG-force data.}
\label{fig:evaluation-generalization-subject}
\end{figure*}

\paragraph{Evaluation metrics}
To assess the performance of our CNN model in estimating fingertip forces, we adopted three quantitative metrics: (1) Classification Accuracy; (2) Normalized Root Mean Squared Error (NRMSE); (3) Coefficient of Determination, $R^{2}$. The model's performance in determining whether a finger exerts force or not at a particular time frame is evaluated by the classification metric, and we only count the model's predictions for a time frame as correct if all five fingers are correctly classified. Using the same notations from~\Cref{sec:method-model}, we have:
\begin{equation}
    \text{NRMSE} = \frac{1}{F_{\max}} \sqrt{\frac{1}{T \cdot I} \sum_{t=1}^{T} \sum_{i=1}^{I} (F_{i}^{t}-\hat{F}_{i}^{t})^{2}}
\end{equation}
\begin{equation}
    R^{2} = 1 - \frac{\sum_{t=1}^{T} \sum_{i=1}^{I}(F_{i}^{t}-\hat{F}_{i}^{t})^{2}}{\sum_{t=1}^{T} \sum_{i=1}^{I}(F_{i}^{t}-\bar{F}_{i}^{t})^{2}}
\end{equation}
where $\bar{F}_{i}^{t}$ gives the mean of $F_{i}^{t}$.


\subsection{Performance of Decoding Finger-Wise Forces}
\label{sec:evaluation-learning}

\paragraph{Experimental setup}
Before considering how our scheme applies to specific or new users, we first evaluate the performance of our model in a user-independent setting, where a single model is trained and shared by all users who contributed data. In particular, the entire training set was used to optimize the model against the joint loss defined in~\Cref{eq:total-loss} for $30$ epochs. An Adam optimizer~\cite{kingma2015adam} with constant learning rate of $1e-4$, $\beta_{1}=0.9$, and $\beta_{2}=0.999$ was adopted. A weight decay factor of $1e-4$ was enforced to mitigate over-fitting. As a post-processing step, we applied a Gaussian filter of window size 10 to the sequence of predicted force values for temporal smoothing. For ablation purposes, we also trained the model using regular $L_{1}$ or $L_{2}$ regression loss only, and cut off predicted force values below a small pre-defined threshold. We used PyTorch~\cite{paszke2019pytorch} to implement all our models as well as to perform training and evaluation.

\begin{table}[t!]
\centering
\caption{Performance comparison with regular regression losses $L_{1}$ and $L_{2}$ as well as the no-smoothing variant of our method.}
 \begin{tabular}{c c c c}
 \hline
 Metric & $L_{1}/L_{2}$ regression & No Smoothing & Ours \\
 \hline
 Accuracy & 85.68\% / 85.12\% & 88.83\% & \textbf{88.83\%} \\ 
 NRMSE & 4.56\% / 4.34\% & 4.02\% & \textbf{3.29\%} \\ 
 $R^{2}$ & 81.89\% / 82.21\% & 83.59\% & \textbf{85.82\%} \\ 
 \hline
 \end{tabular}
 \label{tab:evaluation-ablation-loss}
\end{table}

\paragraph{Results}
The action-wise and subject-wise performance of the user-independent model is summarized in~\Cref{fig:evaluation-learning}. The overall accuracy, NRMSE, and $R^{2}$ are 88.83$\pm$6.13\%, 3.29$\pm$1.76\%, and 85.82$\pm$14.96\%, respectively. On the action side, the model has the highest performance for ring finger pressing, with 92.47$\pm$3.89\% accuracy, 2.10$\pm$0.99\% NRMSE, 90.84$\pm$6.53\% $R^{2}$, and the lowest for index finger pressing, with 85.21$\pm$11.37\% accuracy, 4.12$\pm$3.04\% NRMSE, 70.99$\pm$34.95\% $R^{2}$. On the subject side, one-way repeated measures ANOVA gives $F_{1,8}=1.28$, $p=0.26$ for accuracy, $F_{1,8}=1.18$, $p=0.32$ for NRMSE, and $F_{1,8}=0.82$, $p=0.59$ for $R^{2}$, indicating minor utility discrepancy among subjects. Furthermore, the results of our ablation study are shown in~\Cref{tab:evaluation-ablation-loss}, validating the effectiveness of the proposed joint classification-regression loss and temporal smoothing.
    
\paragraph{Discussion}
The results above demonstrate the feasibility of accurately decoding finger-wise forces from forearm EMG signals and building robust predictive models that can be shared by multiple users. The statistical significance also suggests that the proposed scheme has the potential of being extended beyond an experimental setting and to more general application scenarios. In addition, it is worth noting that such performance holds under the existence of real-world challenges, such as variations across users in sensor positioning, forearm muscle size, forearm hair thickness, etc. The model is resilient to various discrepancies among users, capable of capturing generalizable EMG-to-force patterns, and achieves utility fairness for users, as evidenced by the ANOVA analysis above, all the while maintaining favorable overall performance. 

While these results are statistically rewarding, a remarkably large amount of data is required from each user to support satisfactory performance in practice. Specifically, each participant contributed 22 EMG-Force joint sequences to the training set for the experiment above, which amount to 11 minutes of data. Consider also that there are other inevitable preparations, such as device setup, session break, data pre-processing, and model training. Such delay may become a roadblock for many VR/AR applications in practice. To deploy our neural interface in consumer-level applications, more time-efficient training is essential. This aspect will be addressed in~\Cref{sec:evaluation-generalization}. 


\begin{table}[t!]
\centering
\caption{Performance comparison with vision-based methods~\cite{fallahinia2020comparison,fallahinia2021real,fallahinia2021feasibility} in terms of NRMSE.}
 \begin{tabular}{c c c c c}
 \hline
 Method & Index & Middle & Ring & Mean \\
 \hline
 \cite{fallahinia2020comparison} & 6.1\% & 5.3\% & 10.1\% & 6.2\% \\
 \cite{fallahinia2021real} & 6.1\% & 5.4\% & 9.0\% & 5.9\% \\
 \cite{fallahinia2021feasibility} & 5.7\% & 4.2\% & 8.2\% & 4.9\% \\
 Ours & \textbf{4.7\%} & \textbf{3.7\%} & \textbf{2.4\%} & \textbf{3.7\%} \\
 \hline
 \end{tabular}
 \label{tab:evaluation-comparison}
\end{table}

\subsection{Comparison with Vision-Based Methods}
\label{sec:evaluation-comparison}

Prior works in the literature have explored vision-based solutions to body force estimation, such as inferring contact forces from the dynamics of hand-object interactions using RGB videos~\cite{pham2015towards,zhu2016inferring,pham2017hand,hwang2017inferring,ehsani2020use} and predicting finger-level forces from the color changes in fingernail imaging~\cite{sun2008predicting,grieve20103d,grieve20153,grieve2015optimizing,fallahinia2020comparison,fallahinia2021feasibility,fallahinia2021real}. Similar to our approach, a major benefit of vision-based solution is to bypass on-hand force sensing units. Among these solutions, those based on fingernail imaging also have the potential of delivering accurate and flexible per-finger force estimation for VR/AR applications involving complex hand-object interactions. 
In this experiment, we compare the accuracy and robustness between our method and three recent vision-based methods~\cite{fallahinia2020comparison,fallahinia2021feasibility,fallahinia2021real}.

\paragraph{Experimental setup}
Due to the challenges of reproducing the identical hardware prototype of data acquisition as in \cite{fallahinia2020comparison,fallahinia2021real,fallahinia2021feasibility}, we evaluated our method under the setting adopted by them and compared with their reported performance metrics. In particular, \cite{fallahinia2020comparison,fallahinia2021real,fallahinia2021feasibility} only considered single-finger grasping actions and evaluated their method using the index, middle, and ring fingers. To accommodate their evaluation setting, we separated out the partition corresponding to these three single-finger actions from our EMG-Force dataset, i.e., 3x30 seconds (three sessions) of time-synchronized EMG and force data for each subject and each of the three fingers. Two randomly selected sessions were used to train a single user-independent model, while the remaining session was used to evaluate the trained model.

\paragraph{Results}
The finger-wise and overall performance of force estimation, as evaluated by NRMSE, is shown in~\Cref{tab:evaluation-comparison}. Our method outperforms the three vision-based baselines by $2.5\%$, $2.2\%$, and $1.2\%$ NRMSE on average, respectively. The advantage is especially noticeable in performance for the ring finger, with our method's NRMSE being less than a quarter of~\cite{fallahinia2020comparison} and a third of~\cite{fallahinia2021real,fallahinia2021feasibility}.

\begin{figure}
\centering
  \includegraphics[width=0.9\linewidth]{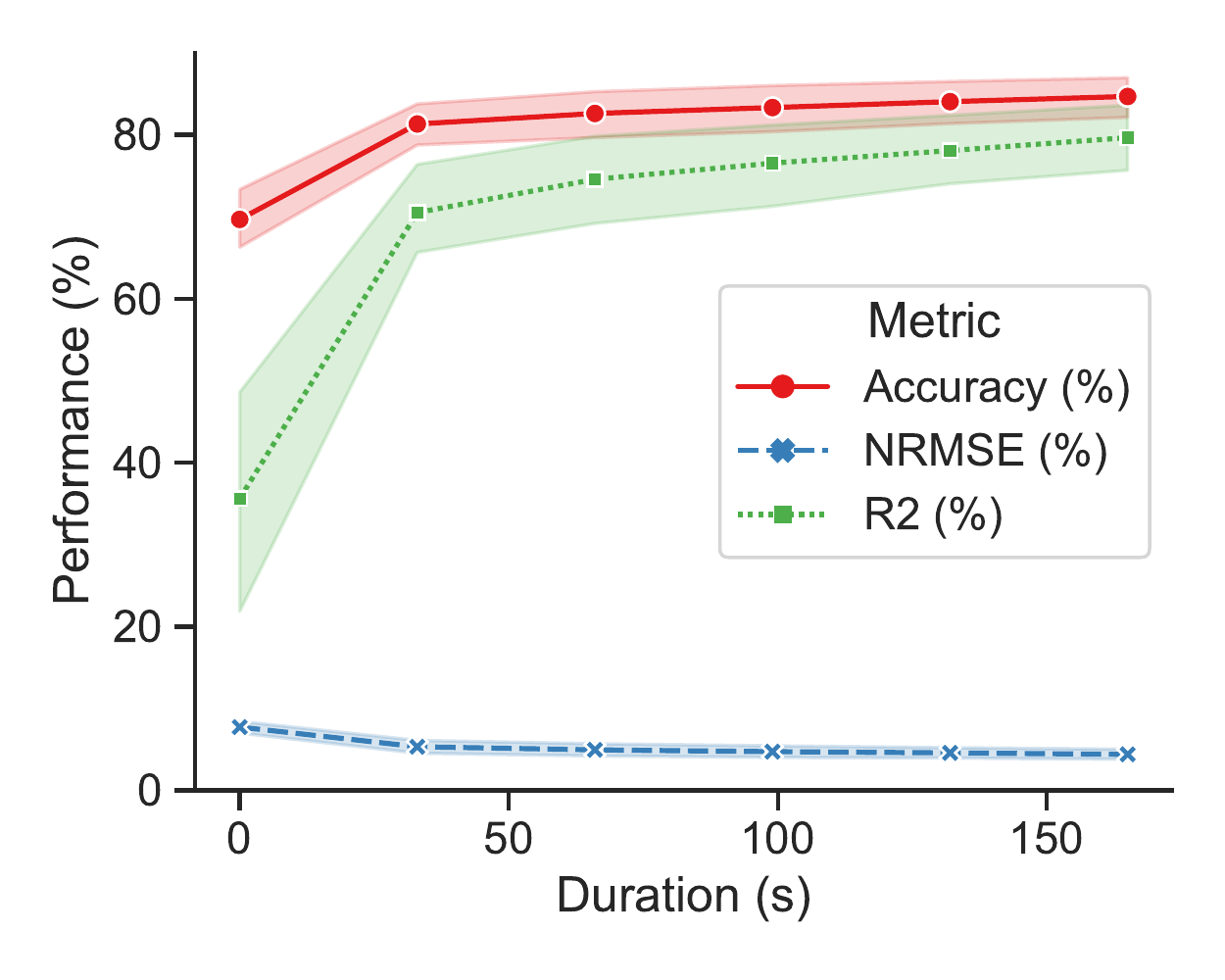}
\Caption{Data-efficient calibration of user-specific model via transfer learning.}
{The performance of user-specific model, as measured by classification accuracy, NRMSE and $R^{2}$, is plotted in function of the amount (i.e., time duration) of data collected from the new user for transfer learning. The translucent band around each curve gives the 95\% confidence interval.}
\label{fig:evaluation-generalization}
\end{figure}

\paragraph{Discussion}
Compared to our EMG-based solution, which actively decodes finger-level forces from the controlling muscles' activities, vision-based methods rely on passive observations and are thus highly sensitive to the variations in external factors, such as ambient occlusions, viewing angles, and lighting conditions. Such degrading effects are more problematic for applications involving complex hand-object interactions or real-wild scenarios. On the contrary, our EMG-based solution has shown high robustness to such mentioned issues by its nature. Besides the vulnerability to environmental factors, fingernail-imaging-based methods are also limited in their functional force range, since the variations in fingernail color get less and less detectable as the force intensity increases. The typical functional range for these methods, as evaluated in~\cite{fallahinia2020comparison,fallahinia2021feasibility,fallahinia2021real}, is around 10N. By contrast, our EMG-based solution is more scalable in terms of force intensity and can robustly estimate forces up to 30N.


\subsection{Individualization and Generalization}
\label{sec:evaluation-generalization}

The analysis of user-independent training in \Cref{sec:evaluation-learning} reveals the need for \emph{time-efficient generalization}. In this section, we investigate how this goal is achievable by extending a pre-trained model for new users using only minimal amounts of data from them. This procedure is commonly referred to as calibration or individualization in human research. Calibration is crucial for machine learning on EMG data since large natural variations exist among different people's muscle-to-EMG patterns. Such discrepancies lead to the well-known generalization challenge that a machine-learned model trained on EMG data commonly fails if directly applied to an unseen user~\cite{phinyomark2018emg}. Therefore, we decide to perform transfer learning to reduce the data requirements for deploying our model to a new user while maintaining satisfactory prediction performance.

\begin{figure}
\centering
  \includegraphics[width=0.99\linewidth]{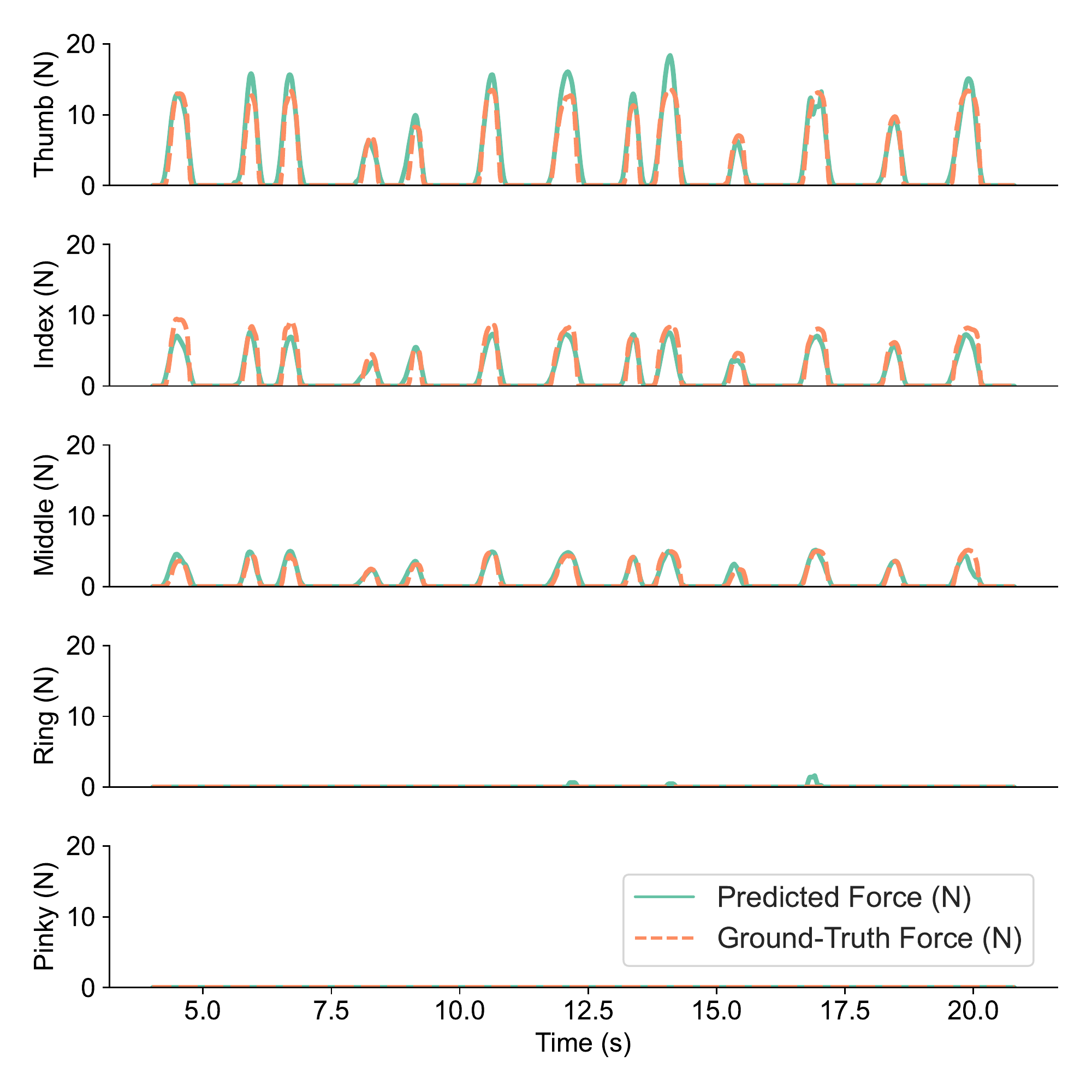}
\Caption{Comparison between predicted finger-wise forces and ground truth.}
{The model is calibrated for subject 8 using only 165-second data from one of his training session and evaluated on a randomly selected sequence from his evaluation session.}
\label{fig:evaluation-force}
\end{figure}

\paragraph{Experimental setup}
To evaluate time-efficient generalization via transfer learning, we adopted an experimental setup similar to cross-validation practices for machine learning tasks. Specifically, we first treated subject 1 (S1) as the new user and optimized the model using two sessions of data from each of the other eight subjects. Next, we fine-tuned the resulting model using a portion of data randomly selected from S1's first session, varying from 10\% to 50\%, to calibrate it into a user-specific model dedicated to S1. Note that 10\% session corresponds to 33-second data. Adam optimizer~\cite{kingma2015adam} with constant learning rate of $5e-5$, $\beta_{1}=0.9$, and $\beta_{2}=0.999$ was adopted. A weight decay factor of $1e-4$ was enforced to mitigate over-fitting. We cycled through the role of new user with each subject to complete the experiment.

\paragraph{Results}
The performance of a user-specific model transferred using 165-second data (50\% of each subject's first session), as measured by classification accuracy, NRMSE, and $R^{2}$, is summarized in~\Cref{fig:evaluation-generalization-subject}. At least 81.23\% accuracy was consistently observed for all subjects' calibrated models except S6, whose model showed 77.77\% accuracy. S1/S4's models achieved over 89.54\% accuracy, surpassing the overall performance of the user-independent model with much less training data. The NRMSE metric revealed larger gaps across subjects, which ranged from 2.80\% to 5.91\%. $R^{2}$ is mostly above 75\%, with the exception of S6's model yielding 67.73\%.

To investigate the minimal amount of data required for calibrating the model towards reasonable utility in practical applications, we analyzed the trade-off between data volume for calibration and resulting model's performance.~\Cref{fig:evaluation-generalization} visualizes the calibrated models' average performance gain as a function of EMG-Force sequences' total length in time. All subjects' models demonstrated rapid improvements as the calibration kicked off and attained 83.33$\pm$4.47\% accuracy, 4.74$\pm$1.09\% NRMSE, and 76.55$\pm$7.88\% $R^{2}$ with $66$-second data only. The growth rate of mean performance then slowed down, and accuracy/NRMSE gradually plateaued when the duration of data exceeded 150 seconds.
Remarkably, these results verify the data efficiency of transfer-learning-based individualization, as evidenced by various accuracy metrics. Taking S8's calibrated model (using 99-second data) as an example, we show a visual comparison between model-predicted and hardware-sensed force values for a randomly selected EMG sequence from their evaluation session in~\Cref{fig:evaluation-force}. Despite the variations in force intensity and temporal spacing, predicted force values generally aligned well with the ground truth, except for a few slight false positives for the ring finger.

\paragraph{Discussion}
Exploiting transfer learning techniques, we demonstrated the feasibility of effectively adapting existing models into dedicated ones for previously unseen users using very limited data from them. 
Further, as indicated by the above analysis and visualized in \Cref{fig:evaluation-generalization}, the high-precision generalization is achieved with simple and rapid (less than 2 minutes) calibration for novel users.

These findings also circle back to our goal of improving physicality for VR/AR environments in two significant ways. First, our system can harness users' natural abilities and predilections for manipulating things they encounter, thereby expanding the space for developers to create VR/AR experiences that map to real-world scenarios and user behaviors. Importantly, we show that this is achievable for any user, with minimal retooling.
Second, as shown in the following section, our system is responsive with low latency. This is crucial, in particular, if we consider that the visual components of AR/VR now routinely refresh at 90 Hz. Any system designed to communicate bodily forces with the virtual environment needs to be agile in timing and nimble in response to data.


\subsection{System Performance}
\label{sec:evaluation-system}

Thanks to the moderate scale and bottleneck design of our CNN model, our system's storage and computing requirements are minor for modern PC hardware. The compact model only contains 1.26M 32-bit floating point parameters ($\simeq$5MB memory). For the EMG sequence streamed at a particular time frame, i.e., 1248 eight-channel EMG samples, our model only generates around 29.19M Multiply-Accumulate Operations (MACs). Note that while our model requires an EMG sequence of 1248 samples (624ms) as input, these data are only retrieved from history to make predictions for the current time step. When applied in practice, our system's latency performance has two important interfering dimensions: 1) \textit{runtime speed}, i.e., the time needed to complete the force predictions on an EMG sequence of 1248 samples; 2) \textit{reaction latency}, i.e., the duration between EMG data generation and force prediction.

As an implementation detail, we performed GPU parallelization with two consecutive EMG sequences that differ by 32 samples and achieved $\simeq$1.2-1.4ms inference time using a GTX TITAN XP GPU, thus obtaining an approximate 0.7ms latency (i.e. over 1000FPS). Note that this result characterizes the system itself rather than the force output, which is also determined by the spectrogram frame rate. With this setting, the system shall wait for both sequences to arrive, introducing an additional 16ms (32 EMG samples) idle time. That is, although GPU parallelization accelerates runtime speed, it also introduces extra reaction latency. Together with the wireless EMG data transmission latency (2ms) and model inference (0.7ms), our system achieves an overall reaction latency of $\simeq$18.7ms, sufficient for most real-time VR/AR applications~\cite{dangxiao2019haptic}.


\subsection{Psychophysical Study: Enhancing Material Perception in Virtual Environments}
\label{sec:evaluation-user-study}

\begin{figure}
\centering
\subfloat[Elastic sheet with \forceCondition (top) and \positionCondition (bottom).]{
    \includegraphics[width=0.99\linewidth]{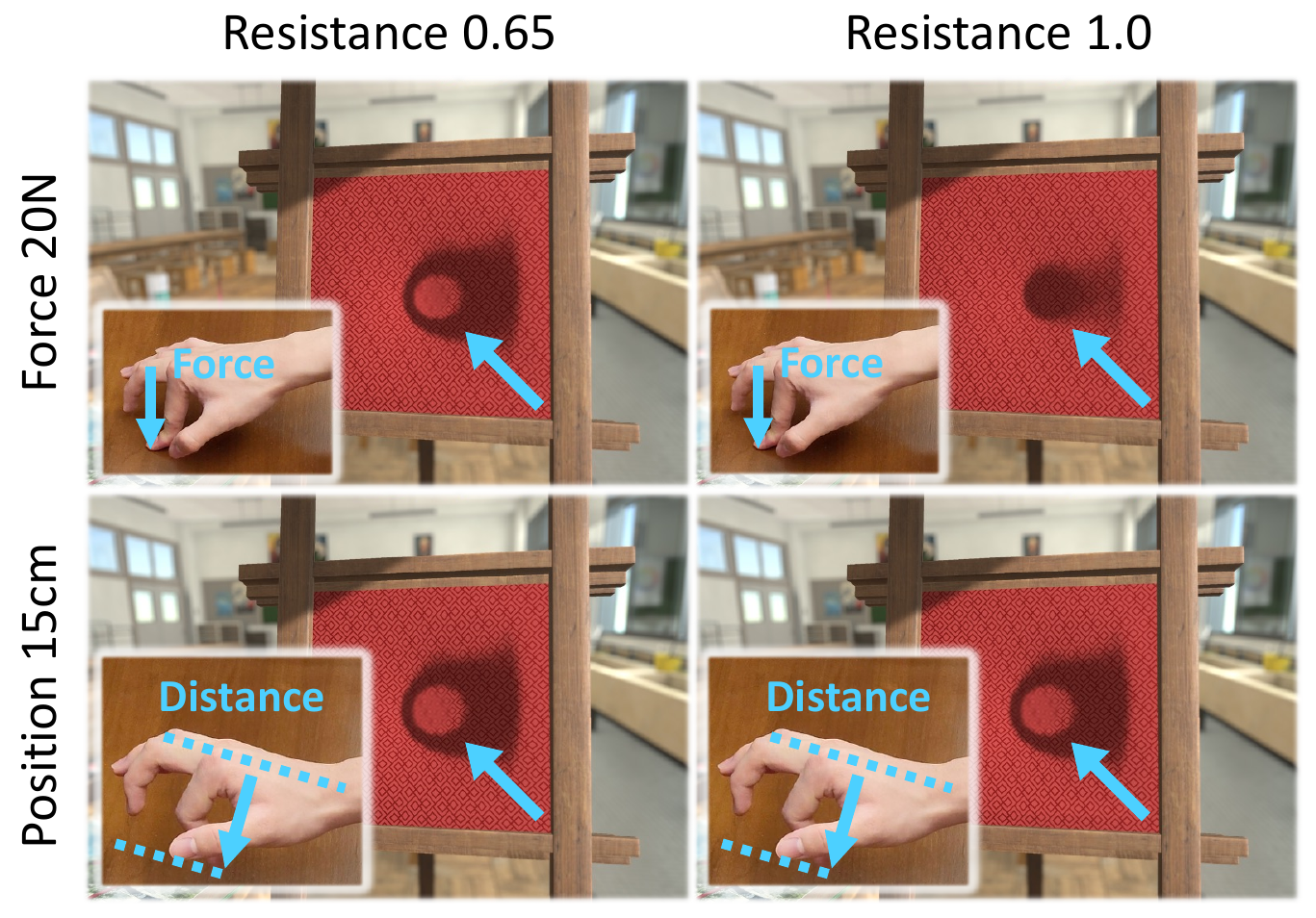}
    \label{fig:evaluation-study-stimuli-sheet}
    } \\
\subfloat[Elastic rod with \forceCondition (top) and \positionCondition (bottom).]{
    \includegraphics[width=0.99\linewidth]{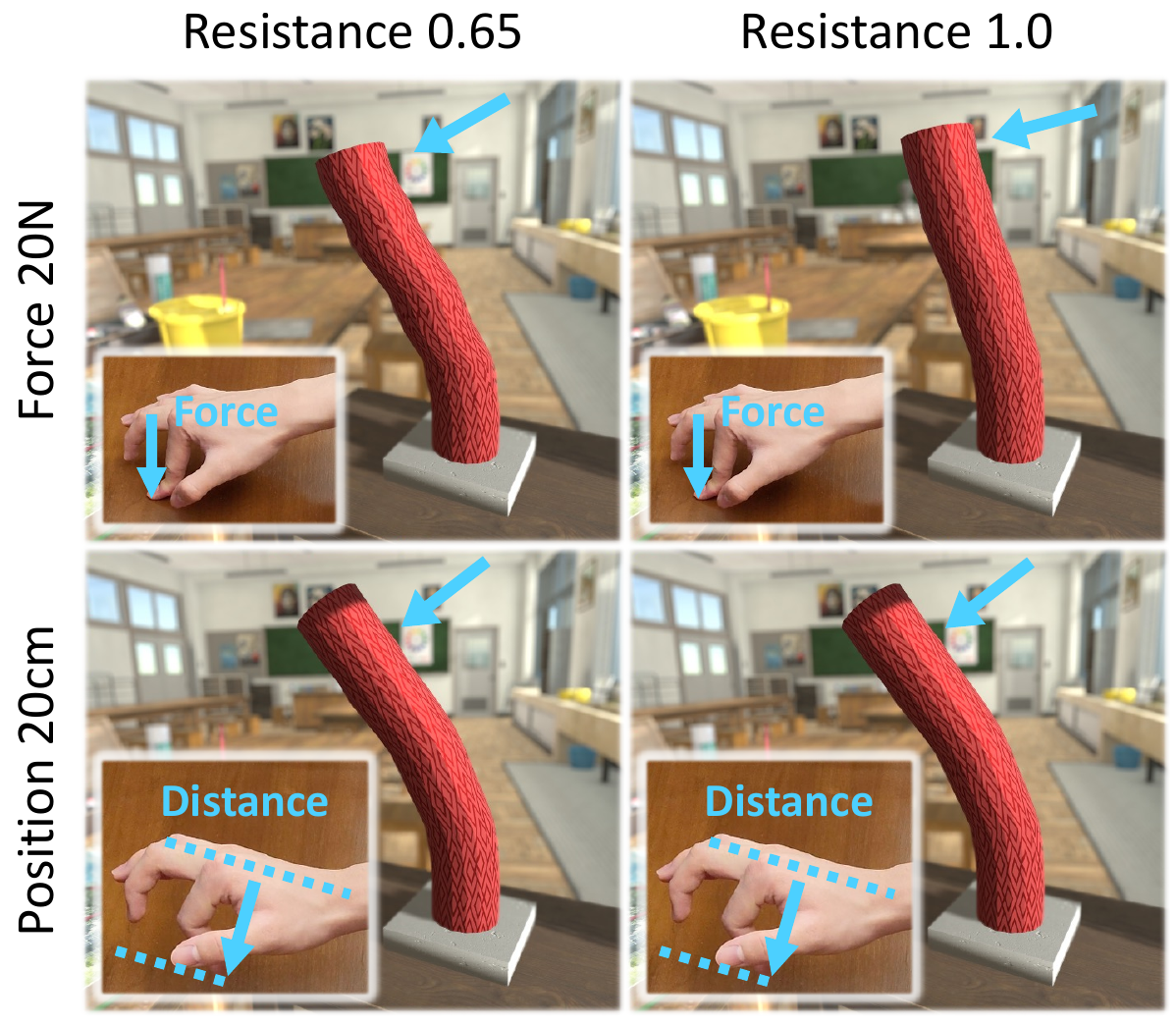}
    \label{fig:evaluation-study-stimuli-rod}
    }
\Caption{Visualization of the stimuli used in our psychophysical experiment.}
{\subref{fig:evaluation-study-stimuli-sheet} illustrates the reaction of two identically-looking elastic sheets differing in material stiffness to inputs from \forceCondition (top) and \positionCondition (bottom). Compared with \positionCondition, the two objects exhibit more natural and prominent difference given the same inputs from \forceCondition.~\subref{fig:evaluation-study-stimuli-rod} shows similar results for an elastic rod. 3D asset credits to SbbUtutuya and Virtual Method at Unity.}
\label{fig:evaluation-study-stimuli}
\end{figure}

A key aim of VR/AR is to create virtual experiences for users as if they were in a physical environment. When interacting with objects in the physical world, we perceive their material properties, such as elasticity and stiffness, through a combination of haptic and visual feedback~\cite{baumgartner2013visual}. To recreate such perceptual realism in virtual environments, it is essential to precisely drive virtual objects' motions and deformations via users' muscular forces. We hypothesize that interfaces with such capability may significantly enhance human perception of virtual objects' material properties. In this study, we evaluate to what extent our system, as a real-time force-aware interface, advances toward this goal.

\paragraph{Participants, setup, and calibration}
We recruited 12 subjects (ages 20-35, 6 female) to participate in the study. A calibration procedure was performed for each subject before starting the experiment by collecting 1 minute of EMG-Force data from him/her to customize the pre-trained user-independent model (as described in~\Cref{sec:evaluation-generalization}). This calibrated model was then used to estimate finger-wise forces on the EMG signals sensed from that subject in real-time. Estimated force values were communicated to a Unity program via the ZeroMQ library. This pipeline allows for direct application of estimated forces to virtual objects in real-time via physical simulation. During the study, the subjects, wearing an Oculus Quest 2 head-mounted display, remained seated and were free to observe a virtual scene. They interacted with virtual objects in their field of view through unconstrained movements of their forearms, hands, and fingers.

\paragraph{Stimuli}
As shown in~\Cref{fig:evaluation-study-stimuli}, the visual stimuli were two geometric primitives (elastic sheet and rod) that are soft and deformable. To enable real-time softbody simulation for low-latency interaction on portable VR headsets, we employed an efficient XPBD~\cite{muller2007position,macklin2016xpbd} implementation by \emph{Virtual Method Studio}~\cite{obi} and only used low-resolution particle models of the virtual objects. It should be noted that simulators' efficiency is orthogonal to the accuracy of model-predicted muscular forces. Therefore, our model can be readily incorporated into \emph{any} simulation system. We adopted deformation resistance of elastic materials as the proxy to represent stiffness with the range $[0,1]$. Deformation resistance measures a physical material's ability to resist externally loaded forces. In this study, we aim to identify participants' discriminative thresholds of virtual objects' stiffness under varying conditions. To avoid visual cues biasing the results, all objects were rendered with identical material and texture, regardless of their physical properties.

\paragraph{Conditions}
For evaluation purposes, all subjects were instructed to employ two interaction methods in sequence during the study (the order was random). Besides our data-driven system for force-enabled interaction (\forceCondition), we also included position-based interaction (\positionCondition) for comparison. Specifically, \positionCondition is a commonly adopted solution in commercial VR/AR systems that lacks force information. It allows users to modify virtual objects’ position, orientation, and shape by colliding their hands with the objects. For \forceCondition, users interacted with virtual objects by manipulating physical proxies while our system estimated their muscular forces. These forces were then used to deform the two virtual objects, including indenting the sheet's center and bending the rod's top. For \positionCondition, we leverage the hand tracking capability of Oculus Quest 2 to estimate the flexion level of users' index finger (distance from the index fingertip to the palm plane) and use it as input for the identical interaction as \forceCondition. Specifically, this value determines the indentation depth for the sheet's center and the bending level for the rod's top.~\Cref{fig:evaluation-study-stimuli} illustrates these interactions. Notably, while physical proxies that resemble the virtual objects will enhance users' experience with \forceCondition, \positionCondition does not benefit from them. To avoid users favoring \forceCondition due to irrelevant features of the physical proxies, we intentionally used a hard and flat table.

\begin{figure}
\centering
\subfloat[Discrimination threshold for \forceCondition and \positionCondition.]{
    \includegraphics[width=0.8\linewidth]{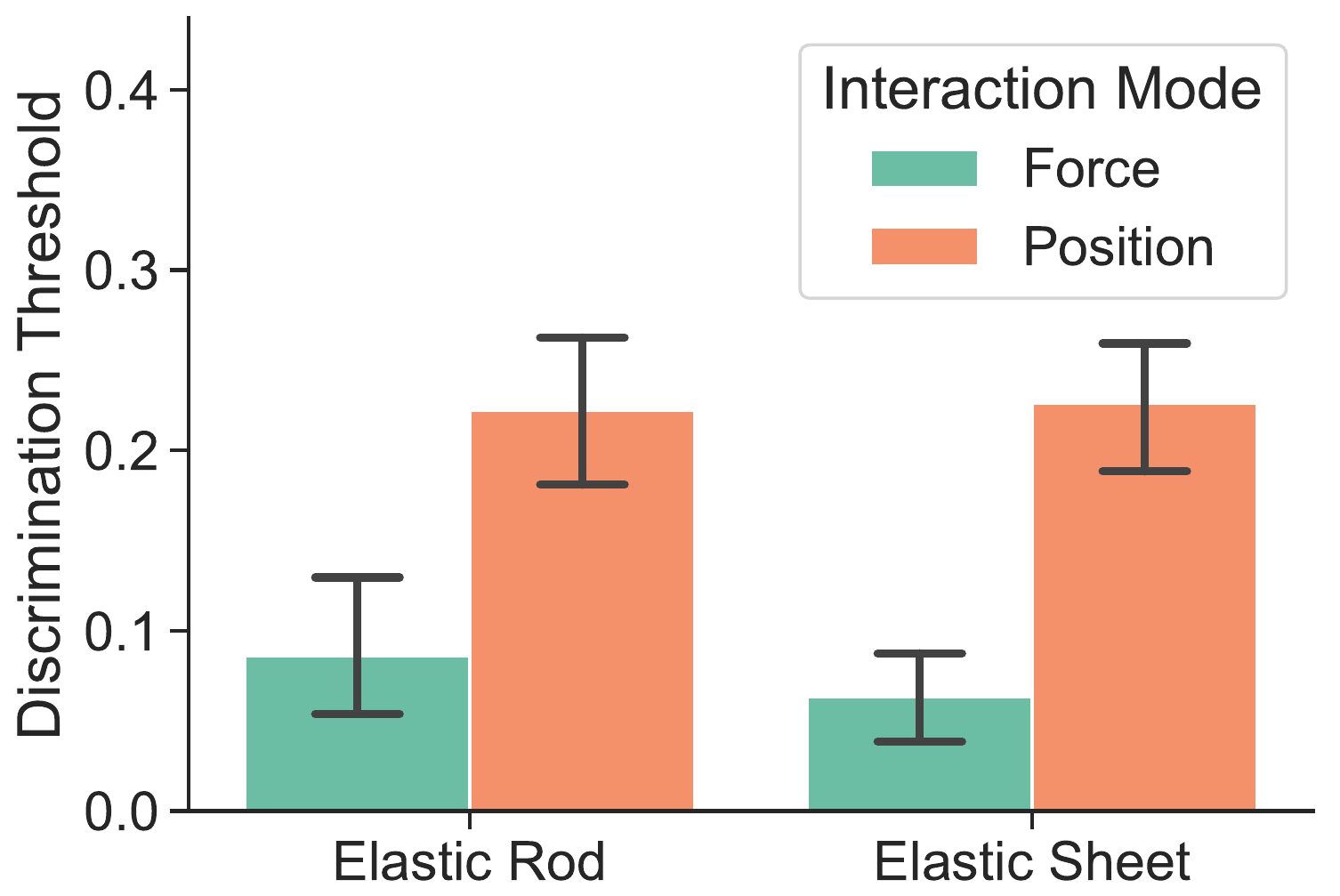}
    \label{fig:evaluation-study-analysis:dist}
    } \\
\subfloat[Discrimination threshold along the decision process of subject 8.]{
    \includegraphics[width=0.99\linewidth]{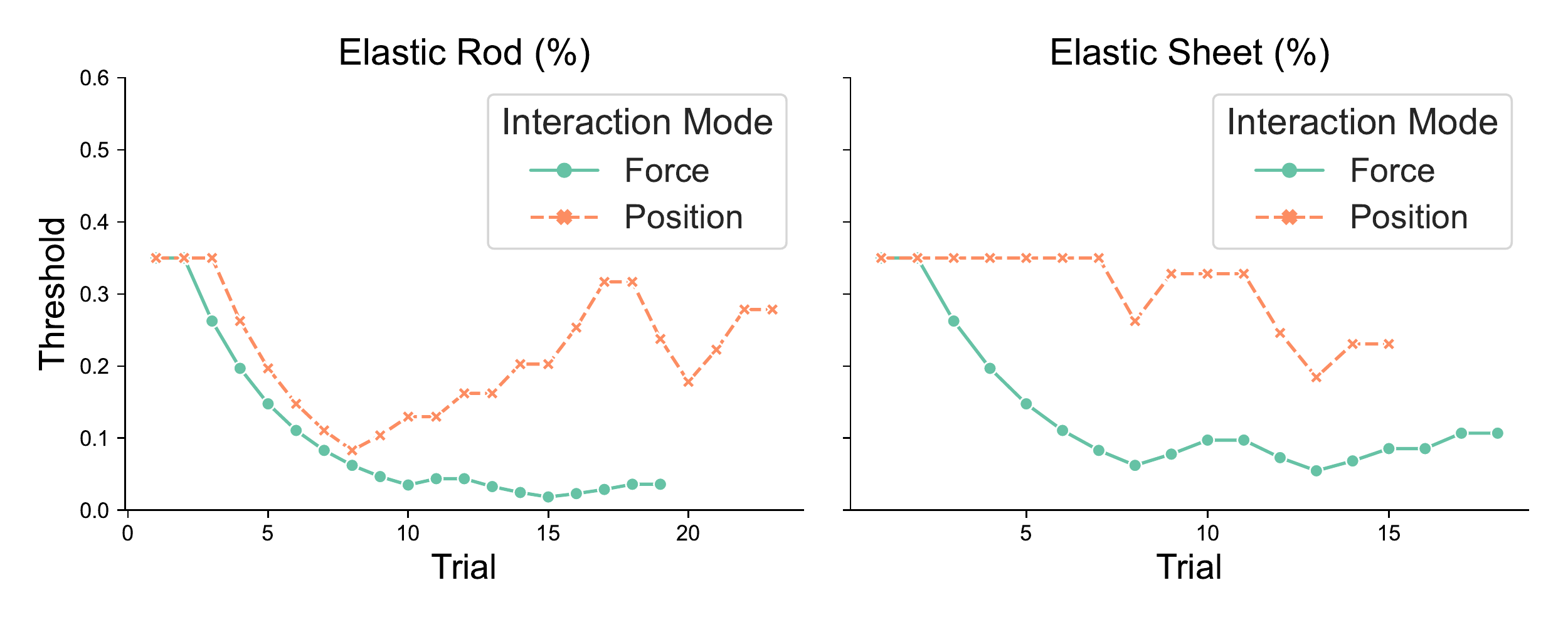}
    \label{fig:evaluation-study-analysis:subA}
    } \\
\subfloat[Discrimination threshold along the decision process of subject 10.]{
    \includegraphics[width=0.99\linewidth]{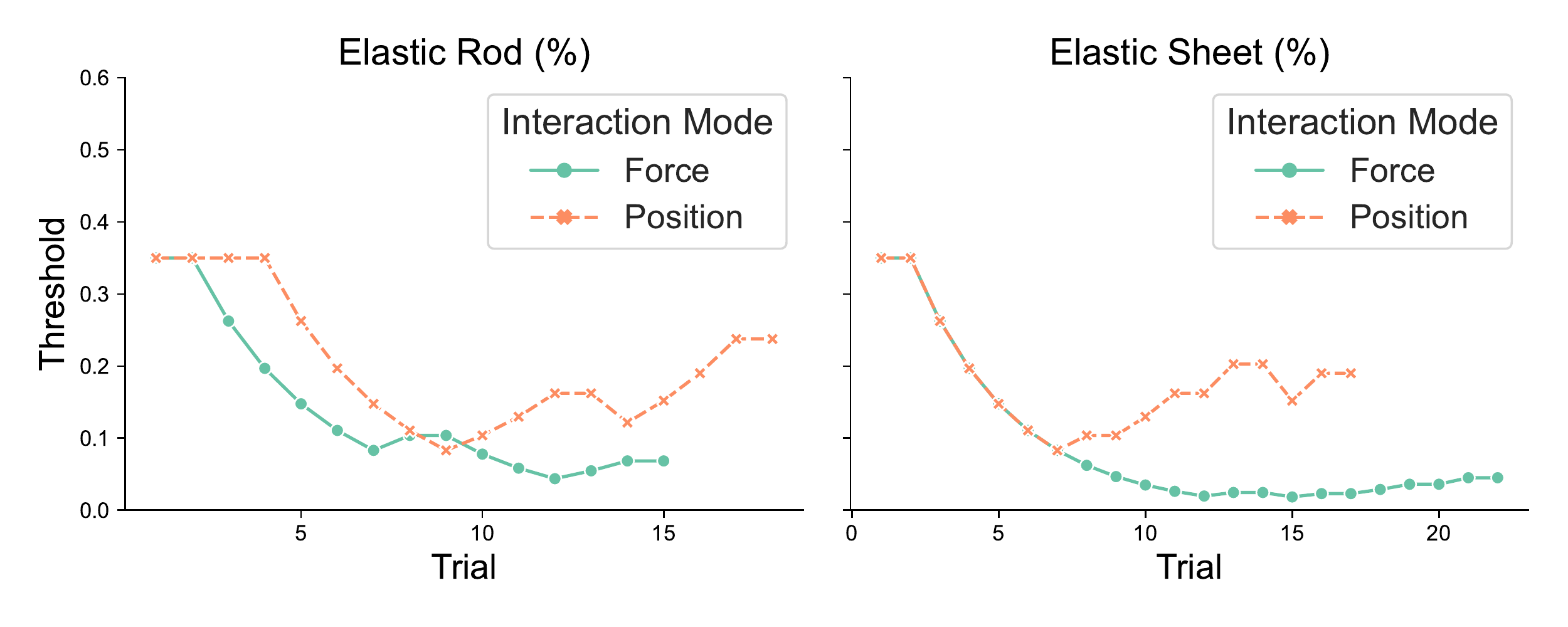}
    \label{fig:evaluation-study-analysis:subB}
    }
\Caption{Psychophysical study on material perception in VR.}
{\subref{fig:evaluation-study-analysis:dist} shows 12 subjects' discrimination thresholds on two geometric primitives for \forceCondition and \positionCondition. The error bars indicate 95\% confidence intervals. A remarkably and consistently lower threshold for \forceCondition can be observed.~\subref{fig:evaluation-study-analysis:subA} and~\subref{fig:evaluation-study-analysis:subB} visualize the discrimination threshold along the decision process of two subjects when they were engaging in our 1-up-2-down staircase protocol with 2AFC trials (which object is stiffer). Note the consistent decreasing trend of \forceCondition, indicating high confidence level when the subject made 2AFC decisions.}
\label{fig:evaluation-study-analysis}
\end{figure}

\paragraph{Task}
To measure participants' (perceptual) discriminative threshold of material stiffness, we employed a psychophysical task as a 2-alternative-forced-choice (2AFC) with a 1-up-2-down staircase procedure (5-reversal to confirm convergence thus termination). For softer and stiffer objects, the experiment started with deformation resistance equal to $0.65$ and $1.0$, i.e. a threshold of 0.35. Each time the threshold got updated, it was incremented or decremented by a quarter of the current threshold, and the two deformation resistance values were updated such that their mean was unchanged. Specifically, the two interaction conditions (\forceCondition/\positionCondition) were sequentially presented to the user for consideration (with a random and counter-balanced order). During the experiment, the participants were instructed to freely interact with the corresponding stimuli and then indicate (using a keyboard) which one of the two stimuli appeared stiffer. They observed the deformation pattern along with the proactive intervention. After each trial, the participants chose one of the two stimuli that appeared stiffer. Each 2AFC trial took 5 seconds. A warm-up session was first performed to allow each individual user to familiarize with the stimuli and the interaction design. For each participant, the entire experiment took about half an hour. The number of trials (ranging from 25 to 38) depended on the speed of the staircase convergence. 
 
\paragraph{Metrics and results}
\Cref{fig:evaluation-study-analysis} visualizes our statistical results. The mean discriminative thresholds of \forceCondition/\positionCondition were 0.09$\pm$0.07, 0.22$\pm$0.07 for the elastic rod, and 0.06$\pm$0.04, 0.23$\pm$0.07 for the elastic sheet, indicating 61.3\% and 72.1\% improvements with \forceCondition, respectively. One-way repeated measures ANOVA shows that the effects of interaction method are statistically significant: $F_{1,11}=20.99$ $p=1.46e^{-4}$ for the elastic rod and $F_{1,11}=46.79$ $p=7.16e^{-7}$ for the elastic sheet.

\paragraph{Discussion}
We designed our psychophysical experiment to test whether users could quantitatively perceive realistic soft objects and their material properties in AR/VR. We used primitive geometries and their natural articulated abilities to explore, examine, and assess them through force-based interaction with the forearm, hand, and fingers. In other words, we wished to test whether users could simply enter a VR/AR scene, start to prod and poke things in that scene, and leave with a sense that the objects responded with realistic physics. 

The results showed a statistically significantly lower discrimination threshold while participants interacted with virtual objects with \forceCondition. That is, the force-visual correlated interaction facilitated significantly more realistic perception of virtual objects' physical characteristics when users were engaged in free manipulation within the virtual world. We regard this as a significant proof of concept for our approach. Consider that, in the real world, humans spend much of their infancy working out how to muster the forces available to them in their arms, hands, and fingers, through ongoing trial and error with the things that they encounter. In essence, we capture the small electrical signals that human muscles cast as they put their skills to use, and we are able to use these signals as indices for machine-learning what that might mean in physics. Asking and answering how our human users believe that the physical response our system yields are realistic-seeming establishes the perceptual foundation of various new possibilities of interfaces.
\section{Applications}
\label{sec:applications}

Beyond enhancing the physical realism of hand-object interactions in VR/AR through more natural and intuitive haptic inputs, our method of decoding forces from EMG signals can also benefit the following \emph{general} application scenarios. We note the following three capabilities---representing virtual humans, virtual interfacing, and virtual control---because they form the ingredients for many \emph{specific} VR/AR applications, e.g., gaming, content creation, assistive technology, training, social communication, etc.

\paragraph{Multimodal data sources for virtual human synthesis}
Synthesizing realistic human behaviors in virtual environments remains an open and essential topic \cite{hassan_samp_2021,yin2021discovering}. While human action data from optical motion capturing systems is common, the knowledge of users' forces on the surrounding physical space that is essential in interactive and dynamic environments \cite{jain2009optimization} remains missing. Our system may provide the data generation foundation that collects the motion-force joint information in daily actions. 

\paragraph{Accessible interfaces}
People with hand/limb impairments may face challenges when interacting with common human-computer interfaces, through actions such as touching and sketching. Bio-prosthetic controls with neural sensors have been emerging as a promising assistive technology \cite{srinivasan2021neural}. The proposed research may predict the intended hand forces by sensing the forearm remotely, allowing for potential applications on assistive interfaces for people with hand impairment or under scenarios such as during driving or cold outdoor temperature.

\paragraph{Ubiquitous control}
The proposed neural interface enables bare-hand interactions and introduces little if any distraction, since it incurs unnoticeable change to the way we operate our hands. Besides, it is usable in most daily situations without safety or privacy concerns. As a result, we can leverage it to replace traditional off-body input devices for ubiquitous control. For instance, we can map patterns of finger pressing to the buttons in a software's control panel or the keys on a musical instrument. As illustrative of broader applications, in the following, we demonstrate that our method can be readily adapted to perform robust finger identification during multi-finger tapping for ubiquitous control.


\subsection{Case Study: Ubiquitous Control via Finger Tapping}
\label{sec:application-control}

Traditional computers typically equip with dedicated control devices such as mouse and keyboard. However, the ultimate goal of VR and AR platforms is a transformative natural and ubiquitous control. To this end, researchers have recently attempted to infer user intention from natural modalities, such as vision~\cite{han2020megatrack,kim2012digits,stearns2018touchcam}, acoustics~\cite{harrison2008scratch,zhang2018fingerping,xu2020earbuddy}, radar~\cite{lien2016soli,wang2016interacting}, and Wi-Fi~\cite{abdelnasser2015wigest,abdelnasser2018ubiquitous}. Despite their support for eyes-free control, these methods are susceptible to environmental interference, such as occlusions and noise, and may suffer from performance decay in complex environments. By contrast, EMG-based solutions electronically tracks users' forearms as input devices. In this experiment, we validate our method's performance while being applied to detect ``click'' actions, which are identified as any finger's tapping.

\begin{figure}
\centering
\subfloat[Multi-finger tapping (right hand).]{
    \includegraphics[width=0.99\linewidth]{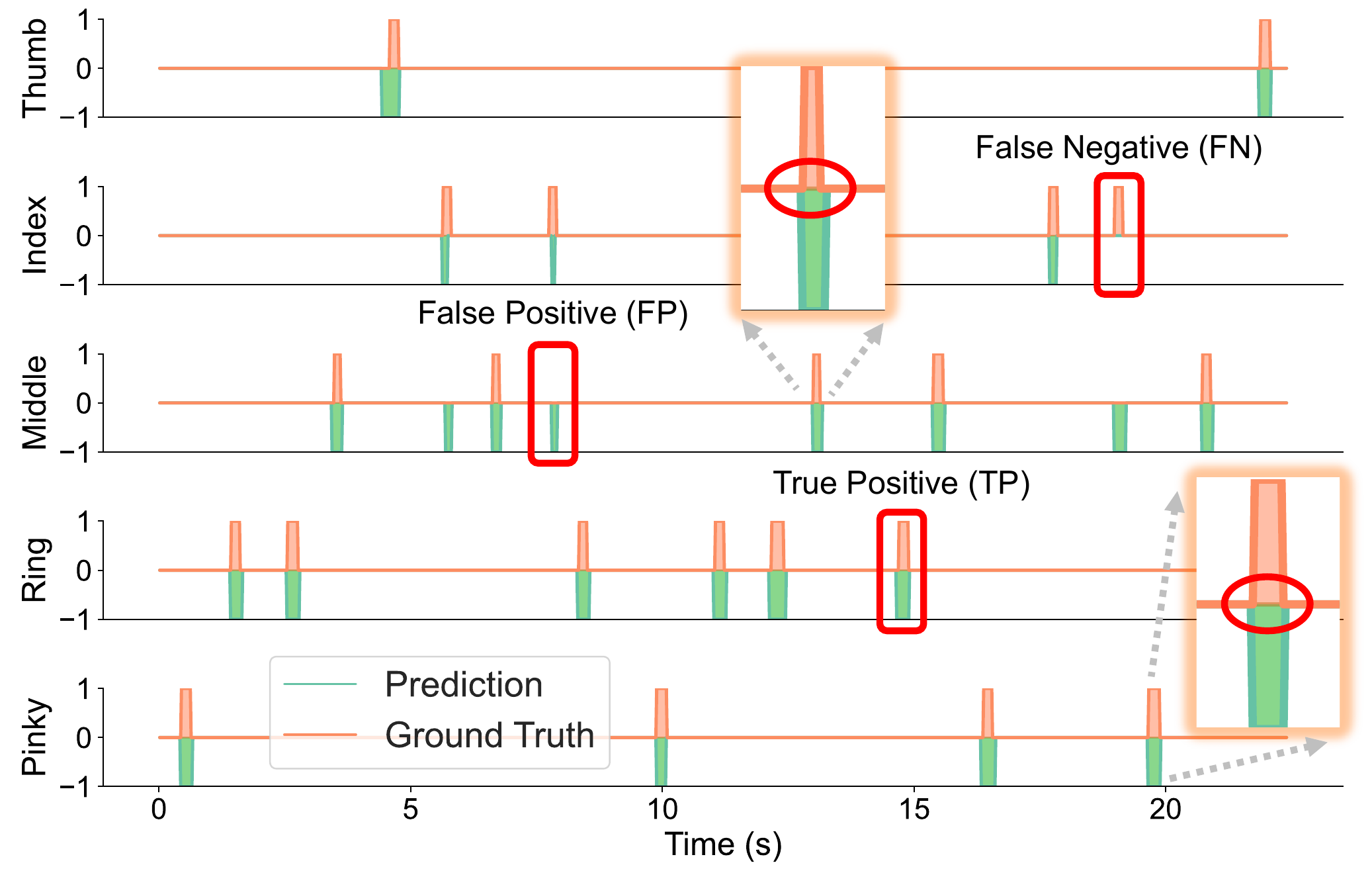}
    \label{fig:evaluation-control-right}
    } \\
\subfloat[Multi-finger tapping (left hand).]{
    \includegraphics[width=0.99\linewidth]{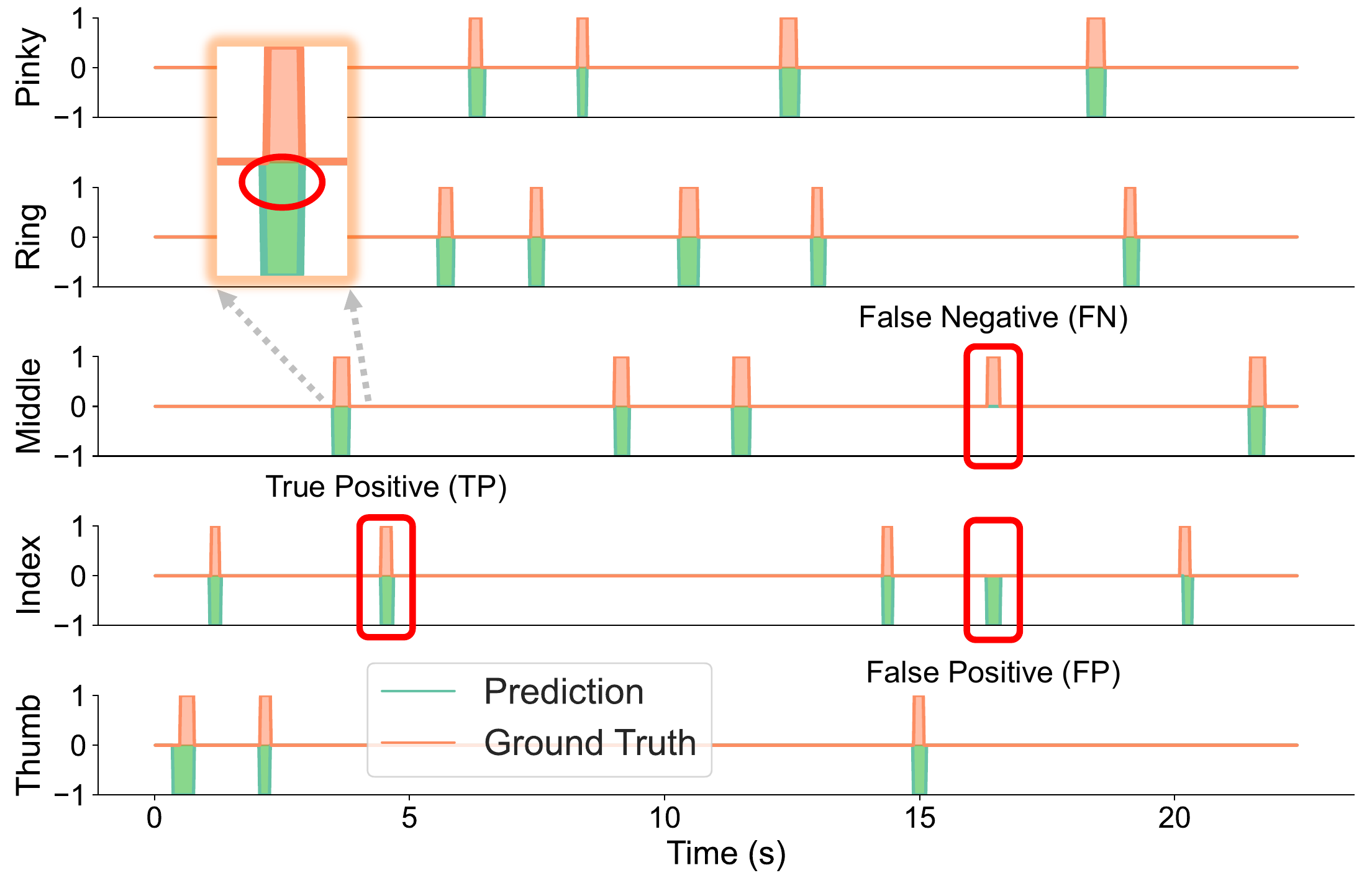}
    \label{fig:evaluation-control-left}
    }
\Caption{Finger identification during multi-finger tapping.}
{\subref{fig:evaluation-control-right} and~\subref{fig:evaluation-control-left} visualize the ground-truth (orange bars) and the identification results by our method (green bars) of two randomly-sampled sequences of multi-finger tapping from the hold-out data. Bar width denotes tapping duration.}
\label{fig:evaluation-control}
\end{figure}

\begin{table}[t!]
\centering
\caption{Tapping finger identification performance.}
 \begin{tabular}{c c c c c c c}
 \hline
 Metric & Thumb & Index & Middle & Ring & Pinky & Mean \\
 \hline
 Precision & 93.3\% & 88.8\% & 84.9\% & 98.4\% & 94.4\% & 92.0\% \\
 Recall & 94.2\% & 78.4\% & 94.7\% & 100.0\% & 96.2\% & 92.7\% \\
 \hline
 \end{tabular}
 \label{tab:evaluation-control}
\end{table}

\paragraph{Experimental setup}
All 10 fingers from both hands are included. One male subject participated in the study. Following the data collection and pre-processing pipeline (\Cref{sec:method-data-collection}), we conducted four collection sessions. During each of the first three sessions, the subject performed single-finger tapping actions and captured $30$ seconds of tapping data for each of the 10 fingers. During the last session, the subject performed random multi-finger tapping actions and captured $150$ seconds of tapping data for each hand. The subject was instructed to tap a trackpad in a natural unconstrained manner in all four sessions. In total, the subject contributed $1200$-second of time-synchronized EMG and force data. The force data was subsequently converted into $\{0, 1\}$ labels, i.e., ``tap'' and ``no-tap''. Two randomly selected sessions (out of the first three sessions) were used to construct the training set. The remaining two sessions were withheld and only used for evaluation. The CNN model was trained using Adam optimizer and cross-entropy loss for the task of per-frame finger-wise ``tap'' or ``no-tap'' classification for $20$ epochs. The learning rate started at $1e-3$ and dropped to $1e-4$ at epoch 10. A weight decay factor of $1e-4$ was enforced to mitigate over-fitting. As a post-processing step, we applied a mean filter of window size 10 to the sequence of predicted tapping probabilities for temporal smoothing. The predicted tapping probability for each time frame was compared with a threshold of 0.3 to determine if it is ``tap''.

\paragraph{Metrics}
We adapted two evaluation metrics from machine learning, precision (P) and recall (R), to accommodate our experimental setting. Here, P/R denotes the proportion of correctly detected tapping among all detected/ground-truth tapping. Note that both metrics are evaluated in a finger-wise manner.
\begin{equation}
    \text{P}_{i} = \frac{\text{TP}_{i}}{\text{TP}_{i}+\text{FP}_{i}}; \quad \text{R}_{i} = \frac{\text{TP}_{i}}{\text{TP}_{i}+\text{FN}_{i}}.
\end{equation}
Here, TP (true positive), FP (false positive), and FN (false negative) denote the number of correctly detected tapping, falsely detected tapping, and undetected ground-truth tapping, respectively. Fingers are indexed by $i \in \{1,2,3,4,5\}$. In particular, a sequence of consecutive ``tap'' predictions of duration longer than 0.1 second is considered as a detected tapping. We consider correct detection if the temporal Intersection over Union (IoU) between its interval and the ground-truth is greater than 0.5, and falsely detected otherwise.

\paragraph{Results}
As shown in~\Cref{tab:evaluation-control}, our CNN model achieves mean precision of $92.0\%$ and mean recall of $92.7\%$ using only 10-minute data for training. The detection quality for the ring finger is the best, with $98.4\%$ precision and $100.0\%$ recall. The index finger showed lower recall $78.4\%$, while the middle finger showed lower precision $84.9\%$. \Cref{fig:evaluation-control} visualizes the ground-truth (orange) and model-detected tappings (green) for two randomly-sampled sequences of multi-finger tapping from the last session, one for each hand. As we can see, most tapping actions are correctly detected by the model, with few false positives and false negatives. Accurate tapping duration and detection latency can also be observed.  

\paragraph{Discussion}
The above results demonstrate our method's applicability to effectively detect finger tapping as an ubiquitous interface. The definitions of precision and recall inherently establish a trade-off between the two metrics: the higher a model achieves in terms of recall, the lower it gets for precision, and vice-versa. Such trade-off can be translated into the context of this specific task as: the more sensitive the model is in detecting finger tapping (more positive predictions), the more actual finger tapping made by the user it will detect (higher recall), and the more mistakes it will make (lower precision). Conveniently, we can increase or decrease the detection threshold for ``tap'' to prioritize over precision or recall, depending on the demand in the actual application scenario. For instance, if we are considering an application where the accuracy of control signals outweighs the response rate, we might consider lowering the model's sensitivity to sacrifice a bit of recall for better precision.
\section{Limitations and Future Work}
\label{sec:discussion}

In this paper, by leveraging EMG sensors on the forearm, we demonstrate the possibility of tracking, predicting, and transferring human muscular forces in the physical world to interactions in virtual representations and environments. In other words, we open-up pathways for the virtual world to react precisely to human-induced physical forces from the real world. Our objective measurements and subject psychophysical experiments support the framework's robustness, accuracy, generalizability, and real-world benefits in enhancing human perceptual understanding of physical materials. This is achieved by our tailored dataset and real-time deep learning approach on muscular signals.

However, several limitations remain for future investigation. First, the dataset that trains our system (\Cref{fig:evaluation-dataset}) were generated by user actions with an off-the-shelf pad-like force sensor. Consequently, the method does not robustly encode complex hand or full-body poses with higher dimensionality and degree-of-freedom, as would be the case if a user was exerting force on more complicated three-dimensional objects. This could be resolved by incorporating data using recent advancements in wearable sensing devices \cite{sundaram2019learning,luo2021learning}, which could enable broader data coverage and thus free-form interaction with complex geometric shapes. Second, our method is practically generalizable but still requires a short (1 minute) calibration process to ensure high-quality predictions. An exciting future direction is extending the framework with unsupervised learning. We believe an automated individualization mechanism may unlock the potential of a fully adaptive framework for arbitrary users without access to the calibration setup. Third, the muscle signals may show different patterns with active (e.g., clenching fist) and resisting (directly interacting with physical objects) forces. Integrating hand tracking and arm pose data into the model may shed light on differentiating the two means. Lastly, to enable accessibility applications, we plan to extend the data and evaluate the system's performance on a larger population, including people with limb impairments. Indeed the ability to sense hand and finger forces and actions directly from forearm muscle signals could establish VR/AR as an entirely new modality for democratizing access to computer graphics applications across a much broader range of interaction abilities. This, we consider, is where development of force-aware VR/AR could be fantastically useful.
\begin{acks}
    This research is partially supported by the National Science Foundation (NSF) under Grant Nos. 2232817, 2225861, 2027652, and 1729815.
\end{acks}


\bibliographystyle{ACM-Reference-Format}
\bibliography{paper.bib}
\appendix

\normalsize
\end{document}